\documentstyle[floats,epsf,aps]{revtex}
\begin{document}
\begin{center}
{\Large \bf Simulated Annealing Clusterization Algorithm for Studying the
Multifragmentation.
}
\end{center}
\vspace*{0.5cm}
\begin{center}
{\large Rajeev K. Puri\footnote{Permanent address: Physics Department,
Panjab University, Chandigarh-160 014, India} and Joerg Aichelin} 
\\
~~
\vspace*{0.2cm}\\
{\it SUBATECH, Laboratoire de Physique Subatomique et des
Technologies Associees\\ UMR Universite de Nantes, IN2P3/CNRS,
Ecole des Mines de Nantes\\ 4 rue Alfred Kastler,
F-44070 Nantes, France.
}
\vspace*{0.5cm}
\end{center}
\baselineskip 24pt
 We here present the details of the numerical realization of the recently
 advanced algorithm developed to identify the fragmentation 
 in heavy ion reactions. This new algorithm is based on the Simulated
 Annealing method and is dubbed as Simulated Annealing Clusterization Algorithm
 [SACA].  We discuss the different parameters used in the Simulated
 Annealing method and present an economical set of the parameters which is
 based on the extensive analysis carried out for the central and peripheral
 collisions of Au-Au, Nb-Nb and Pb-Pb. These parameters are crucial for
 the success of the algorithm. Our set of optimized 
 parameters gives the same results as the most conservative choice 
 , but is very fast. We also discuss the nucleon and
 fragment exchange processes which are  very important 
 for the energy minimization and finally
 present the analysis of the reaction dynamics using the new algorithm.
 This algorithm is can be applied whenever one wants to 
 identify which of a given number of
 constituents form bound objects. 
\newpage
\begin{center}
{\huge I. Introduction:}
\end{center}
\vspace*{0.2cm}
\baselineskip 24pt

     In recent years, a lot of efforts  has been made (in experiments and
     theory) at low, intermediate and relativistic energies to understand
     the physics  which drives heavy ion reactions.
     A new generation of electronic devices made it 
     possible to  measure a multitude of observables  at the same time 
     which can give 
     information about the hot and dense nuclear matter formed during a 
     reaction \cite{Tsang93}. 
     In a heavy ion reaction the density can be as high as 
     2-4 times the normal nuclear matter and one may reach  
     temperatures of about 100 MeV \cite{Puri94}.
     The properties of the nuclear matter at high densities are not only of 
     importance
     for nuclear physics, but are also of great use for the astrophysical
     studies especially for supernova studies. 
     Unfortunately, there is no method to  measure directly the  properties 
     of hot and compressed
     nuclear matter formed during a reaction \cite{Tsang93}. What one can
     observe are single hadrons. Their properties are mostly determined
     in the late stage of the expansion and it is quite difficult to find
     observables sensitive to the early stage. For this 
      one has to rely 
     on the theoretical (simulation) models.
     One can simulate the reaction from the start to
     the end where we find cold nuclear matter in the form of nucleons,
     light and heavy fragments \cite{Aich91}.
     The most important information which
     one would like to extract from the simulation are the time scales of
     different phenomena. One would like to know, for example, when  particles
     are  created, when fragments are formed, whether 
     they carry any information
     about the hot and compress phase etc.? The key question associated with
     the time scale of the fragment emission is whether it is a thermal or
     a dynamical process i.e. whether the fragments are created after the
     system has thermalized or can already be recognized early, before
     a possible thermalization sets in . This would point to initial-final
     state correlations \cite{Puri96, Goss97, Dors93, Sune98}. In addition,
     several conjectures on the equation of state, especially those in which
     the nuclear interaction is strongly momentum dependent, could not be 
     tested so far in simulations because the nuclei become unstable. Here
     an early fragment recognization would allow to study these equations of
     state.
     \\

     We shall concentrate here on multifragmentation. 
     All theoretical models used to study heavy ion collisions
      are based on the transport nucleons and mesons only.
      Therefore, for the study of multifragmentation 
      a method has  to be deviced to group
     the nucleons into free nucleons and fragments. In the past, one has
     taken the spatial correlations among nucleons to group them into
     fragments \cite{Aich91}.
     Naturally, this approach cannot detect different fragments 
     which are (almost) overlapping and therefore 
     will give a single big fragment 
     during the early stage of the reaction where density is quite high.
     In other words, simple coordinate space  
     approaches cannot address the question of the time scale of fragment
     formation.
     To study that one
     needs a method where fragments can be identified even if they are  
     overlapping, i.e. methods which are based on phase space.\\

     We conjecture that in nature at any given moment of the reaction that
     configuration is realized which gives the largest binding energy.
     That this concept is meaningful and gives sensitive results will be
     demonstrated later. 
     To find the most bound configuration we  are confronted with two problems.
     \\
     a) The huge number of possible configurations.\\
     b) The fact that the number of entities changes. Whereas the number
     of nucleons is constant, the number of free nucleons and fragments
     is a variable. The problem caused by this fact will be discussed later.\\

     One may approach this problem by simple iterative methods.
     They, however, do not garranty that a global minimum is obtained but
     may arrive at a local minimum \cite{garc93}.
     First attempt to overcome this problem has been
     advanced in ref. \cite{Dors93}.
     Though this method works fine for small systems,
     its simple numerical implementation 
     poses serious problems for studying the heavy systems where the number
     of different configurations increases tremendously and almost
     always the algorithm remained stuck in 
     a local minima. 
     To deal with  the more interesting large systems a
     sophisticated algorithm is needed 
     which can handle the huge number of different
     configurations and finds the  configuration with
     maximal binding energy in a reasonable
     amount of computational time. In addition, it should be able to 
     overcome any type of local minima.\\

     We here present the details and technical aspects of such a new
     algorithm which is based on the simulated annealing method and is
     quite general in nature. The question addressed here requires an
     numerical approach which is not specific to  the problem described.
     Apart from multifragmentation 
     the energy  minimization is needed,
     for example, in the 
     nuclear structure calculations, in cluster
     radioactivity, in hadron physics etc.
     In nuclear cluster radioactivity, one is interested in the relative 
     yields of different fragments which are emitted by the decaying nucleus.
     There, one assumes the isotopic  distribution and the 
     energy minimization is needed
     to find the most bound isobaric distribution \cite{Gupt93}.
     \\

     Naturally, before one can talk about  cluster formation, 
     one first needs the phase space coordinates of
     the particles.
     We here use the Quantum Molecular Dynamics (QMD) model 
     \cite{Aich91} 
     (as an
     event generator) to generate the time evolution of the 
     phase space coordinates of  nucleons in a nuclear reaction.
\\

     Our paper is organized as follow: 
     The section II deals with the short description
     of the QMD model and a detailed description of the algorithm . 
     The numerical realization of the  algorithm is presented in
     section III and  we summarize the results in section IV.  
\vspace{0.2cm}
\begin{center}
{\huge II. The Formalism}
\end{center}
\vspace*{0.2cm}
\baselineskip 24 pt

We here summarize shortly the {\bf Q}uantum 
{\bf M}olecular {\bf D}ynamics [QMD] model and  then give the details of our
new algorithm designed for multifragmentation. 
For more details on the  QMD approach, we refer the reader 
to \cite{Aich91}.

%
\begin{center}
{\Large (i) The QMD approach:}
\end{center}
      The QMD model is based on molecular dynamics and hence is an
n-body theory which simulates the heavy ion reactions between ${30 A
\cdot MeV}$ to ${1A \cdot GeV}$ on an event by event basis \cite{Hart98}.
Here each nucleus is represented by a coherent state of the form ( $ \hbar $
=1 )
\begin{equation}
\phi_{\alpha} (x_1,t) = \left( \frac{2}{\pi L} \right)^{3/4}
e^{- \left( \vec x_1 - \vec x_{\alpha} - \frac{\vec p_{\alpha}t}{m}\right)^2
/2L}
~~e^{i\vec p_{\alpha} \left( \vec x_1-\vec x_{\alpha} \right) }
~~e^{-\frac{i p_{\alpha}^2
t}{2m}}. \label{e1}
\end{equation}
The wave function has two time dependent parameter $x_{\alpha}$, $p_{\alpha}$.
We fix the Gaussian width $(L)$ to 1.08 fm$^3$. In QMD calculations,  
nucleon $\alpha$ moves on a quasi-classical trajectory as obtained by a
variation solution of the n-body Schrodinger equation:
\begin{equation}
\dot{\vec{x}_{\alpha}} = \frac{p_\alpha}{m} + \nabla_{p_{\alpha}} \sum_{\beta}
<V_{\alpha \beta} (x_{\alpha}, x_{\beta}, p_\alpha, p_\beta)>, \label{e8}
\end{equation}
\begin{equation}
\dot{\vec{p}_{\alpha}} = -\nabla_{\vec{x}_{\alpha}} \sum_{\beta}
<V_{\alpha \beta} (x_{\alpha}, x_{\beta}, p_\alpha, p_\beta)>. \label{e9}
\end{equation}
Here $p_{\alpha}$ and x$_{\alpha}$ are the centriods of the 
Gaussian wave functions
in momentum and coordinate space which represent the nucleon $\alpha$.
 The potential has the form \cite{Aich91}
\begin{equation}
<V_{\alpha \beta} (x_\alpha, x_\beta)> 
= \int{d^3}{x_1} {d^3}{x_2} <\phi_{\alpha}
\phi_{\beta} \mid V(x_1, x_2)| \phi_{\alpha} \phi_{\beta}>. \label{e10}
\end{equation}
In addition, the nucleons interact via stochastic elastic and inelastic
NN collisions.
  In principle, our approach to find the fragments 
  is independent of the algorithm which
  generates the phase space coordinates. Therefore, QMD may be
 replaced by any other  model
 ( like simple molecular dynamics model \cite{Dors93}, 
  Boltzmann-Uhling-Uhlenbeck model etc. \cite{Baur85})
  which is able to generate 
  the phase space coordinates of the particles.
  Due to its n-body nature, the QMD model is more
  appropriate to study the fragment formation in heavy ion collisions
  than one body models. 
 \\

\vspace*{0.2cm} 
%
\vspace*{0.2cm}

\begin{center}
{\Large (ii) A Survey of Heavy Ion Reaction:} 

\end{center}

 During the simulation of the reaction, we store 
 the phase space coordinates of all nucleons at several time steps. 
 As the QMD model simulates the time evolution of nucleons
 , the stored phase space distribution is that of nucleons only.
 Our basic assumption is that in nature that configuration is realized
  which gives the largest binding energy.
  Therefore, a method has to be adopted 
 to  group the nucleons in free nucleons and fragments.
 The nucleons  within a fragment will be 
 bound by some binding energy. 
 In a very simple model,
 one could  consider the nucleons being a part of the same fragment if their
 centriods are closer
 than some spatial distance $r_{max}$.
 This model is called minimum spanning tree
 [MST] method \cite{Aich91, Sune98} . One generally takes $ 2 \le r_{max}
  \le 4$.
 By definition, this method
 cannot address the fragment distribution 
 during the violent phase of the reaction where whole nuclear matter
 is compressed and is confined to few fermis.
 The MST method at this time will
 give one single large fragment. 
 More disturbing, 
 the fragments  detected by the MST method 
 can contain nucleons with very large relative momenta. These fragments will
 be unstable and will decay after a while by fissioning or by 
 emitting nucleons. To improve the model, 
 a cut in momentum space 
 has been also suggested recently by one of us and collaborators \cite{Sune98}.
 This cut (which limits the maximal allowed relative momentum of two 
 nucleons in the same fragment) is quite effective
 in central collisions
 where most of the fragments are created during a reaction, but 
 has no effect on the fragment
 distribution in peripheral 
 collisions where the fragments are  produced due to the  
 decay of the spectator matter.\\ 
 
 If one combines the cuts in momentum and in coordinate space to
 a binding energy cut,
 one sees that  several groups of nucleons are indeed not fragments, but a 
 group of unbound nucleons which are close in spatial space. One
 has to  follow the  reaction for a long time until  this group of nucleons
 decays in 
 light and heavy fragments
 which are well separated in the  coordinate space and can be detected with
  the standard MST algorithm. 
 The critical time  is generally assumed to be about 300 fm/c.\\

To give the reader a more clear picture, we simulated the reaction 
Au-Au at 600
MeV/nucl. and at impact parameters of b = 3 and 8 fm, respectively, and display
 some key quantities in fig.1.
 The solid and dotted lines 
represent the reaction at 8 and 3 fm, respectively. 
The first row shows the evolution of
mean density and of the collisions  as a function of time.
As expected, a higher  density and
collision number can be seen 
in central collisions compared to peripheral collisions.
One also notices that the high reaction rate terminates at about 40-60
fm/c. Afterwards, we observe only collisions of nucleons in the same
fragment. 
The second row shows
the evolution of spectator ( filled circle) and participant ( filled
triangle) nucleons.  A participant nucleon is defined as  a nucleon which has
undergone at least one collision. One sees that in central 
collisions 99$\%$ of nucleons
have experienced a collision until 40 fm/c. As a results the directed 
transverse flow saturates as early as 40 -60 fm/c.
The fig. 1(e) displays the evolution of the size of the 
largest fragment $A^{max}$ detected
by the normal MST method with $ r_{max}$ = 
4 fm. We see 
one big fragment (consisting of 394 nucleons) at the time
when the density is high.  
After
about 120 fm/c we are able to find the "stable" fragment,  which still
decreases in size due to evaporation.
Is this a realistic
identification of the  largest fragment?
To answer this question, we applied a binding energy cut on
the fragments detected in MST method. We first analyze the fragments with
MST method and then pass all the fragments (with mass $\ge$ 3) through
an energy filter which recognizes a fragment only if it has at least a binding
energy of 4 MeV/nucl. Otherwise it considers the MST
fragment as a set of free nucleons. This approach is labelled as MST$^\star$.
 In
both central and peripheral collisions, the largest fragment detected
in MST is not a bound fragment at intermediate times.
One gets properly bound fragment 
after about 120 fm/c after emitting the nucleons what lowers of course
the binding energy.  One should keep in  mind that
in peripheral collisions, one has two big (spectator) fragments and a fireball
at mid-rapidity region without fragment.\\ 

\begin{center}
{\Large (iii) Simulated Annealing Clusterization Algorithm [SACA]:\\
}
\end{center}

Our new approach can be summarized as follows. We assume that : 

\fbox{1} The nucleons from target and projectile are grouped into fragments (of 
any
size) and into free nucleons.

\fbox{2} Though the nucleons inside a fragment can interact with each other,
they do not interact with the nucleons from other fragments or free 
nucleons.

\fbox{3} That pattern of nucleons and fragments is realized in nature
which gives the highest binding energy.

To avoid that at intermediate times too many fragments are assumed
(which finally break apart), we employ in addition a
binding energy check.
In order to form a fragment, the considered group
of nucleons has to have a minimal binding energy given by 
\begin{equation} \label{ebind}
\zeta =  \sum_{\alpha=1}^{N^f} \left [ 
\sqrt{(\vec p_\alpha - \vec P^{cm}_{N^f})^2 + m_\alpha^2} - m_\alpha + 
\frac{1}{2} \sum_{\beta \ne \alpha}^{N^{f}} V_{\alpha \beta}
 (x_\alpha, x_\beta) \right]
< ~~\mbox{$L_{be}$} \times N^f,
\end{equation}
%
with $L_{be}$ =  -4.0 MeV if $N^f$ $\ge$ 3 and $L_{be}$ = 0 otherwise. In 
this equation, $ N^f$ is the number of nucleons in a fragment, 
$\vec P_{N^f}^{cm}$ is the center-of-mass momentum of the fragment. 
The binding energy criteria will make 
sure that no loosely bound fragments are  formed in our approach.
In reality these loosely bound  
fragments are not stable and decay during the reaction. 
The problem is that we have to find the most bound configuration among
a huge number of possible patterns ( composed of nucleons and
fragments). 
In order to cope with this complicated problem, we employ the  
simulated annealing technique and hence this algorithm is dubbed as
"Simulated Annealing Clusterization Algorithm (SACA)".


One is tempted to start the search for the most bound cluster configuration 
by an iterative minimization
method  (also known as neighborhood search or local search). In this method, 
starting from a given configuration a new one
is constructed. The new 
configuration is accepted only if it lowers the binding energy.
The drawback of this procedure is that it may terminate 
at a local minimum. To improve this limitation, 
several modification can be imagined \cite{aarts87}:

1. To execute the algorithm for a large number of the initial configurations.
This will finally allow to reach the global minimum. This is very time
consuming.

2. To use a algorithm which can jump over local
minima and hence one can reach the global minima. This clearly 
depends strongly on the problem. Therefore its applications are limited. 

3: To generalize the iterative method so that the transitions which yields 
 a higher binding energy are always accepted.
 In addition, the transitions which yield
 a lower binding energy are also accepted  with a certain probability.
This algorithm is known as simulated annealing method
\cite{aarts87}. Its name is 
based on the fact that this algorithm is akin to the
one used for cooling the solids.  
The simulated annealing method is a sequence of metropolis algorithms 
\cite{metropolis53} with decreasing control parameter $\vartheta$. The 
control parameter $\vartheta$ can be interpreted 
as a "temperature". For each Metropoliscity at a given temperature, we
perform a sequence of steps until the binding energy does not change anymore.
Each step 
is executed
as follows:

1: Given some initial configuration $\psi$ with energy $\zeta_\psi$, 
a  new configuration $\varphi$ with energy $\zeta_\varphi$ is generated in the  
neighborhood of  $\psi$ using a Monte-Carlo procedure.

2: Let the energy difference between  $\psi$ and $\varphi$ is
  $\Delta \zeta $ = $\zeta_\varphi$ -$\zeta_\psi$.

3.: If $\Delta \zeta$ is negative, the new configuration is always accepted.
If $\Delta \zeta$ is positive,  it is accepted with a probability 
exp$(-\Delta \zeta/\vartheta)$.
At the start, the control parameter $\vartheta$ 
is taken to be large enough for that
most all attempted transitions are accepted. This is to overcome
any kind of the local minima. 
After the binding energy remains constant, a 
gradual decrease in the control parameter 
$\vartheta$ is
made and the Metropolis  algorithm is  repeated. 

Note that there is no change in the coordinates of 
the nucleons. 
 One should also note that employing this method does guaranty in the
limit of infinite steps to reach 
the global minimum. Evidence that one reaches the ground state can
be provided by obtaining the same fragment pattern  for different starting
configurations. \\
 
 To start with, a  random configuration $\psi$ ( 
 which consist of
 fragments and free nucleons) is chosen.
 The total energy associated with  $\psi$
 configuration is given by 

\begin{eqnarray*}
\zeta_\psi = 
	    \sum_{\alpha=1}^{N^f_1}\left\{ 
            \sqrt{(\vec p_\alpha - \vec P^{cm}_{N^f_1})^2 + m_\alpha^2}-m_\alpha
            +\frac{1}{2} \sum_{\beta \ne \alpha}^{N^f_1}
            V_{\alpha \beta} (x_\alpha, x_\beta)\right\}_1 \\
  +\cdots
	    \sum_{\alpha=1}^{N^f_\nu}\left\{ 
            \sqrt{(\vec p_\alpha - \vec P^{cm}_{N^f_\nu})^2 + m_\alpha^2}-m_\alpha
            +\frac{1}{2} \sum_{\beta \ne \alpha}^{N^f_\nu}
            V_{\alpha \beta} (x_\alpha, x_\beta)\right\}_\nu \\
+	    \sum_{\alpha=1}^{N^f_\mu}\left\{ 
            \sqrt{(\vec p_\alpha - \vec P^{cm}_{N^f_\mu})^2 + m_\alpha^2}-m_\alpha
            +\frac{1}{2} \sum_{\beta \ne \alpha}^{N^f_\mu}
            V_{\alpha \beta} (x_\alpha, x_\beta)\right\}_\mu \\
+\cdots
	    \sum_{\alpha=1}^{N^f_n}\left\{ 
            \sqrt{(\vec p_\alpha - \vec P^{cm}_{N^f_n})^2 + m_\alpha^2}-m_\alpha
            +\frac{1}{2} \sum_{\beta \ne \alpha}^{N^f_n}
            V_{\alpha \beta} (x_\alpha, x_\beta)\right\}_n \\
\end{eqnarray*}
Here $N_\mu^f$ is the number of nucleons in a fragment $\mu$,
$\vec P^{cm}_{N^f_\mu}$
is the center of
mass momentum of the fragment $\mu$ and  
$V_{\alpha \beta}(x_\alpha, x_\beta)$ is the
interaction energy  between nucleons $\alpha$
and $ \beta$ in a 
given fragment $\mu$.
Note that the total energy is the sum of the energies of individual fragments
in their respective center of mass system. Therefore, 
$ \zeta_\psi$ differs from the (conserved) total energy of the
 system because (i) the kinetic energies of fragments calculated in
 their center of masses
and 
(ii) the interactions between fragments/free nucleons are neglected.
At present a simple static interaction is implemented, but one can use
 the algorithm for arbitrary interactions.
\\

 A  new configuration is generated using Monte-Carlo procedure 
 by either a)  transferring a nucleon from some
 randomly chosen fragment to another fragment or by b) setting a nucleon
 of a fragment 
 free or c) absorbing  a free nucleon into a fragment. Let 
the new configuration $\varphi$ be generated by transferring a nucleon from 
fragment $\nu$ to fragment $\mu$. Then the energy 
of new configuration $\varphi$ 
is given by:  
\begin{eqnarray*}
\zeta_\varphi= 
	    \sum_{\alpha=1}^{N^f_1}\left\{ 
            \sqrt{(\vec p_\alpha - \vec P^{cm}_{N^f_1})^2 + m_\alpha^2}-m_\alpha
            +\frac{1}{2} \sum_{\beta \ne \alpha}^{N^f_1}
            V_{\alpha \beta} (x_\alpha, x_\beta)\right\}_1 \\
  +\cdots
	    \sum_{\alpha=1}^{N^f_\nu-1}\left\{ 
            \sqrt{(\vec p_\alpha - \vec P^{cm}_{N^f_\nu-1})^2 + m_\alpha^2}-m_\alpha
            +\frac{1}{2} \sum_{\beta \ne \alpha}^{N^f_\nu-1}
            V_{\alpha \beta} (x_\alpha, x_\beta)\right\}_\nu \\
+	    \sum_{\alpha=1}^{N^f_\mu+1}\left\{ 
            \sqrt{(\vec p_\alpha - \vec P^{cm}_{N^f_\mu+1})^2 + m_\alpha^2}-m_\alpha
            +\frac{1}{2} \sum_{\beta \ne \alpha}^{N^f_\mu+1}
            V_{\alpha \beta} (x_\alpha, x_\beta)\right\}_\mu \\
+\cdots
	    \sum_{\alpha=1}^{N^f_n}\left\{ 
            \sqrt{(\vec p_\alpha - \vec P^{cm}_{N^f_n})^2 + m_\alpha^2}-m_\alpha
            +\frac{1}{2} \sum_{\beta \ne \alpha}^{N^f_n}
            V_{\alpha \beta} (x_\alpha, x_\beta)\right\}_n \\
\end{eqnarray*}
Note that in this procedure, the individual energies of all fragments except 
for the donar fragment ($\nu$) and the receptor fragment ($\mu$) 
remain the same. The change in the energy from 
 $\psi$  $\longrightarrow$ $\varphi$ is given by
\begin{equation}
\Delta \zeta = \zeta_{\varphi}- \zeta_\psi.
\end{equation}

Between the Metropolis algorithms, we cool the system 
by decreasing the control parameter $\vartheta$. 
A decrease in the temperature  
means that we narrow the energy difference which is accepted
in a metropolis step.   
 After many Metropolis steps, 
 we should arrive at a minimum i.e. the most bound configuration.
The problem is, however , that we usually arrive at a local minimum only.
Between the local minimum, we find huge maxima. Let us give an example:
Assume we have two fragments , but the most bound configuration  would be 
one single fragment which combines both. Now each exchange of a single nucleons
raises the binding energy and only the exchange of all nucleons at the
same time lowers the total binding energy. This effect is well known in
chemistry, where it is called Activation energy.
In order to avoid this, we add , therefore, a second 
simulated annealing algorithm in which not anymore the nucleons are
considered as the particles  which are exchanged in each Metropolis step 
(like in the first simulated annealing), but the entities ( fragments
 or nucleons) obtained after the first step. This second stage of 
 minimization is called fragment exchange procedure. This fragment exchange
 procedure is capable of overcoming any local minima.\\

Note that
even in this second stage of the minimization, the free nucleons can 
be exchanged as before.
 The total energy associated with any 
 configuration $\Psi$ during second stage of iterations is given by 

\begin{eqnarray*}
\zeta_\Psi = 
	   \left \{ \sum_{\alpha=1}^{N_{S_1}} \left [ 
            \sqrt{(\vec p_\alpha - \vec P^{cm}_{N_{S_1}})^2 
	    + m_\alpha^2 }- m_\alpha
           +\frac{1}{2}  
	   \sum_{\beta \ne \alpha}^{N_{S_1}} V_{\alpha \beta} (x_\alpha,x_\beta) 
	   \right ]\right\}_1 \\
+\cdots
	   \left \{ \sum_{\alpha=1}^{N_{S_\nu}} \left [ 
            \sqrt{(\vec p_\alpha - \vec P^{cm}_{N_{S_\nu}})^2 
	    + m_\alpha^2 }- m_\alpha
           +\frac{1}{2}  
	   \sum_{\beta \ne \alpha}^{N_{S_\nu}}
	   V_{\alpha \beta} (x_\alpha,x_\beta) 
	   \right ]\right\}_\nu \\
+
	   \left \{ \sum_{\alpha=1}^{N_{S_\mu}} \left [ 
            \sqrt{(\vec p_\alpha - \vec P^{cm}_{N_{S_\mu}})^2 + 
	    m_\alpha^2 }- m_\alpha
           +\frac{1}{2}  
	   \sum_{\beta \ne \alpha}^{N_{S_\mu}}
	   V_{\alpha \beta} (x_\alpha,x_\beta) 
	   \right ]\right\}_\mu \\
+\cdots
	   \left \{ \sum_{\alpha=1}^{N_{S_n}} \left [ 
            \sqrt{(\vec p_\alpha - \vec P^{cm}_{N_{S_n}})^2 
	    + m_\alpha^2 }- m_\alpha
           +\frac{1}{2}  
	   \sum_{\beta \ne \alpha}^{N_{S_n}} V_{\alpha \beta} (x_\alpha,x_\beta) 
	   \right ]\right\}_n \\
\end{eqnarray*}
Here $N_{S_\mu} = \sum_{i=1}^{N^f_{S_\mu}}
 N^i_{S_\mu}$is the number of nucleons in a super-fragment $S_\mu$.
$N^i_{S_\mu}$ is the number
of nucleons in the i-th fragment contained in the super-fragment $S_\mu$ and
$N^f_{S_\mu}$ is the number of pre-fragments contained in the super-fragment
$S_\mu$.
The $\vec P^{cm}_{N_{S_\mu}}$ is the center of
mass momentum of the super fragment  $S_\mu$ and  
$V_{\alpha \beta }(x_\alpha, x_\beta)$
is the interaction energy between nucleons $\alpha$ 
and $\beta$ in a 
given super-fragment. Note that now the particle $\alpha$ interacts with its
fellow nucleons in the same pre-fragment and also with the nucleons of other 
pre-fragments
which are contained in a new given super fragment $S_\mu$. 
\\

 Now the new configuration is generated using Monte-Carlo procedure 
 by either a)  transferring a pre-fragment from some
 randomly chosen super-fragment to another super-fragment 
 or by b) setting a pre-fragment 
 free or c) absorbing  a single isolated 
 pre-fragment into a super-fragment. Let us suppose that a 
new configuration $\Phi$ is generated by transferring a pre-fragment {\bf i}
(
with mass $N^i_{S_\nu})$ from 
super-fragment $\nu$ to super-fragment $\mu$. The associated energy 
of new configuration $\Phi$ 
reads as :  
\begin{eqnarray*}
\zeta_\Phi = 
	   \left \{ \sum_{\alpha=1}^{N_{S_1}} \left [ 
            \sqrt{(\vec p_\alpha - \vec P^{cm}_{N_{S_1}})^2 
	    + m_\alpha^2 }- m_\alpha
           +\frac{1}{2} 
	   \sum_{\beta \ne \alpha}^{N_{S_1}}
	   V_{\alpha \beta} (x_\alpha,x_\beta) 
	   \right ]\right\}_1 \\
+\cdots
	   \left \{ \sum_{\alpha=1}^{N_{S_\nu}-N^i_{S_\nu}} \left [ 
            \sqrt{(\vec p_\alpha - \vec P^{cm}_{N_{S_\nu}-N^i_{S_\nu}})^2 
	    + m_\alpha^2 }- m_\alpha
           +\frac{1}{2} 
	   \sum_{\beta \ne \alpha}^{N_{S_\nu}-N^i_{S_\nu}}
	   V_{\alpha \beta} (x_\alpha,x_\beta) 
	   \right ]\right\}_\nu \\
+
	   \left \{ \sum_{\alpha=1}^{N_{S_\mu}+N^i_{S_\nu}} \left [ 
            \sqrt{(\vec p_\alpha - \vec P^{cm}_{N_{S_\mu}+N^i_{S_\nu}})^2 + 
	    m_\alpha^2 }- m_\alpha
           +\frac{1}{2}  
	   \sum_{\beta \ne \alpha}^{N_{S_\mu}+N^i_{S_\nu}}
	   V_{\alpha \beta} (x_\alpha,x_\beta) 
	   \right ]\right\}_\mu \\
+\cdots
	   \left \{ \sum_{\alpha=1}^{N_{S_n}} \left [ 
            \sqrt{(\vec p_\alpha - \vec P^{cm}_{N_{S_n}})^2 
	    + m_\alpha^2 }- m_\alpha
           +\frac{1}{2} 
	   \sum_{\beta \ne \alpha}^{N_{S_n}}
	   V_{\alpha \beta} (x_\alpha,x_\beta) 
	   \right ]\right\}_n \\
\end{eqnarray*}
 The only difference between the particle and the fragment exchange procedure 
 occurs for the bound nucleons. Now the bound nucleons cannot change
 their identity neither by being absorbed nor by becoming free. They will 
 remain bound in a pre-fragment. The pre-fragment
 itself can change its identity
 by either getting transferred to a new super-fragment, or be set free.
As in the first stage, we calculate the energy difference between
the new and the old configurations $\Delta \zeta$ and the metropolis procedure is
continued till the most favored configuration is obtained. 

In summary, the actual procedure is as follows:  We first simulate
 the  nucleus-nucleus collision 
 using the QMD model and store the phase-space coordinates of all
 nucleons at several time steps. At each stored time step, we apply the  
 SACA to find the most bound configuration which consists of
 nucleons and fragments (of any size).  For a faster
 convergence of the algorithm, any cluster decomposition irrespective 
 whether it fulfills the binding energy check (eq. \ref{ebind}) or not is 
considered.
 Therefore, it is likely that several clusters 
 may fail to fulfill eq. (\ref{ebind}). 
At the end of the algorithm when
 the most bound configuration is found, 
 we check the binding energy ( eq. 5) 
 of each superfragment explicitly
 and mark all
 super-fragments violating this condition. The nucleons belonging to an 
 'inhibited' (marked) cluster are further on treated as free nucleons.
 The minimizing procedure of the simulated annealing mechanism 
 is invoked again until a configuration is found where all fragments fulfill
 eq. (\ref{ebind}).
The heavy fragments are usually more bound than the lighter ones. 
We have carried out a detailed analysis and found that
these are always the light fragments ( with masses 3 or 4 ) which at the end
of the iterations are unbound or loosely bound. \\

In the following, we discuss the numerical realization of the algorithm
and present a detailed analysis of the influence of different
parameters 
used in simulated annealing method. 
\\
\begin{center}
{\huge III. Numerical Realization:}
\end{center}
\vspace*{0.2cm}


The simulated annealing
algorithm  has  several parameters to be determined : 
the initial and the final value of the 
control parameter $\vartheta$, the number
of metropolis steps to be executed at a 
given value of control parameter (i. e. length of Markov chain) 
, the decrease of the control parameter and the termination of the algorithm.
This set of parameters is also referred as cooling schedule in the literature
\cite{aarts87}.
One needs to choose the following
parameters explicitly:\\

 1.: The initial value of the control parameter $\vartheta_{i}$. This will be
 referred as temperature.

 2.: The final value of the control parameter $\vartheta_{f}$ [
 i.e. the termination procedure].

 3.: The length of the Markov chain $M_{ch}$.

 4.: A rule to fix the decrement in the control parameter $\sigma$.

 Following \cite{aarts87}, 
 we use a so called simple cooling scheme and present the analysis of 
 our extensive
 tests made for the collisions of Au-Au at 600 MeV/nucl. and at an
 impact parameter of 8 fm. We have also analyzed the results for the
 collisions of Pb-Pb (central) and Nb-Nb (central and peripheral).
 The results of our analysis are
 independent of the masses of the colliding nuclei and 
 also of the impact parameter. For our analysis
 we chose a conservative  
 set of the above parameters and then try to find an 
 optimized set of the parameters
 which yields the shortest
 computational time. 
  We use the following set of parameters if not stated otherwise:\\

  The initial temperature $\vartheta_i$ is taken to be 4 MeV.  
  The length of Markov chain is taken to be 70$\eta$. 
   ;$\eta$ being the
  number of  entities at the start of the minimization.
  After each
  markov chain [ = $ 70 \eta $], the temperature is decreased using a simple
  law :
  $$ \vartheta_{i+1} = \sigma \cdot \vartheta_i, $$
   with $\sigma$ = 0.95. Finally, the algorithm is 
  terminated if there is no change in the binding energy for a large number
  of iterations [ = $60 \eta$]. The details of each of these parameters
  will be presented the following paragraphs. 

 
 \fbox{i} {\it The Initial Configuration}: 
 We have to choose a random initial distribution initially to evoke the  
 simulated annealing minimization. In our procedure, 
 we distribute the nucleons (of the two colliding nuclei) 
 into few cells. The transfer of nucleons is allowed 
 among these cells. Naturally, the final outcome should be
 independent of the number of cells we choose. 
 In fig. 2, we present the 
 outcome of a single QMD event when exposed to SACA with
 a different number of initial cells.
 The displayed reaction is of Au-Au at 600 MeV/nucl. and impact
 parameter of 8 fm. Here we vary the number of 
 cells between  two and 394  ( that
 is by treating each nucleon as a free particle). 
 We see that the variation in the cell number does not 
 affect the final fragment distribution. 
At zero fm/c, the simulated annealing method finds two nuclei 
( i.e. the projectile and target) which shows the 
 validity of the annealing method. One also notices that the 
 binding energy of the system remains constant 
 between 40 fm/c and 200 fm/c.
In other words, the most bound configuration found in SACA at
 40 fm/c and 200 fm/c is  approximately the same.\\

 In fig. 3, we display the  evolution of most bound configuration using
 the two extremes : 2 cells and all particles free at  
 0, 40, 120 and 200 fm/c, respectively. Note that
 the high density phase is reached 
 around 40 fm/c. Between 120 and 200 fm/c, one should not expect 
 much change as the reaction is already finished and there is only some 
 rearrangement of the nucleons in the fragments. 
 In case of 2 cells ( two initial clusters) the  
  algorithm first breaks each of the clusters into large number of
 free nucleons (because free nucleons have zero energy) and some 
 light fragments. After several hundred thousands iterations, it 
 starts rearranging the nucleons into bound fragments. 
 It is interesting to note that after some initial differences,
 the evolution of most bound configuration is quite the same in  both cases.
\\

 As stated in the algorithm section, we choose the new configuration
 $(\varphi$ or $ \Phi$)
 by transferring a nucleon/pre-fragment from one fragment/superfragment to
 another. These fragments are chosen by Monte-Carlo method. It would be 
 interesting to study the effect of different Monte-Carlo procedures 
 (applied in SACA) on the 
 fragment distribution.  The different Monte-Carlo procedures can be generated
 using different random  numbers. 
 In figure 4, we display the
 fragment distribution ( i.e. the largest
 fragment A$^{max}$, the number of free nucleons and of 
 intermediate mass fragments IMF's $  A \ge 5$ ) and the energy of the
 most bound configuration at three different times i.e. at zero fm/c, 40 fm/c and
 200 fm/c, respectively, for the different iterations.
We see, as it should be, a almost complete independence. \\

 \fbox{ii}. {\it The Initial Value $\vartheta_i$}:
 One of the key  features of
 the Metropolis  algorithm is that
 it also accepts the transitions which increases 
 the cost function i.e. the energy. 
 Therefore, the initial value 
 $\vartheta_i$  should be such that the most of the 
 attempted transitions are accepted
 during first iterations.
 In other words, exp$(-\Delta \zeta/\vartheta_i)$ $\sim$ 1. A practical way to 
 implement the sequence of $\vartheta_i$'s
 is given by Johnson et.al. \cite{john87}.
 There the average increase in the energy over large
 number of iterations $\bar{\Delta \zeta}$ is related with $\vartheta_i$ by
\begin{equation}
 \mbox{acceptance ratio} ~~~~~\chi = \mbox{exp} \left\{ -\bar{\Delta\zeta
                    }/\vartheta_i\right\}.
\end{equation}
i.e. 
\begin{equation}
 \vartheta_i = \frac{\bar{\Delta \zeta}}{\mbox{ln} (\chi ^{-1})}.
\end{equation}
Generally, the acceptance ratio $\chi $ should be close to 1.  
The choice of the value 
$\vartheta_i$ (or the temperature) depends very strongly on the problem
 at hand. It should be kept in mind that a very large 
value of $\vartheta_i$ will lead to huge
computational time whereas a very
small value will  lead to less attempts which are accepted by the Metropolis
algorithm and consequently which may lead to  a wrong final distribution.
In fig. 5, we show
the same reaction as reported in fig. 1-4, but at different values of initial
temperature $\vartheta_i$.
Here the other parameters are kept unchanged.
The variation of $\vartheta_i$ between 1 MeV and
500 MeV has no effect on the fragment distribution at zero fm/c. 
We have just two gold nuclei initially. On contrary, one can see 
 some differences at 40 fm/c. 
 A very small value of $\vartheta_i$ 
 ($\leq$ 3-4 MeV) leads to a heavier
 A$^{max}$ (95) compared to the average A$^{max}$ ( $\approx$ 42) and 
 as a result  fewer IMF's and nucleons are emitted. 
 Similar conclusions can be drawn 
 at 200 fm/c. A very low value
 of $\vartheta_i$ apparently freezes the initial 
 configuration. 
 The results are  more stable for $\vartheta_i \ge 3 MeV$. 
 Therefore, we choose the $\vartheta_i$= 5 MeV.\\


In fig.6,  we display the evolution of the most
  bound configuration in Au-Au reactions using $\vartheta_i$
  = 5 MeV and 500 MeV.
  We see that when one iterates the reaction with very large 
  initial temperature (=500 MeV), almost all
  attempted transitions are accepted in the Metropolis algorithm. 
  With a moderate value of the temperature $\vartheta_i$ = 5 MeV,
  only some selected configurations are  
  accepted. The minimization of the energy with $\vartheta_i$= 500
  MeV results in the vibration around 
  the same fragments for very long time. 
  An (unnecessary) large value of $\vartheta_i$ does not help
  to establish an early equilibrium. In contrary, 
  one needs huge computation time (in terms
  of iterations) to find the most bound configuration. The same can be
  achieved with moderate value of $\vartheta_i$ = 5 MeV with far less costs.
\\

\fbox{iii}. {\it The Length of the Markov Chain: M$_{ch}$:~~~}
Here we fix the control parameter $\vartheta$ and 
execute the algorithm for a fixed number of Metropolis steps. 
We construct  a sequence of fragment configurations
Q = $\{ \psi, \varphi, \Psi,....., \Phi\}$. 
One should note that here we have an initial configuration
$\psi$
and new configuration $\varphi (= \psi+1$)
is generated by a random matrix. Thus, 
the number of iterations, and hence the length of Markov chain should be 
long enough to ensure an equilibrium.  
In fig. 7, we show the results with Markov chains of different length.
 The length of the markov chain  $\eta$ is displayed in  the units of the
 total number of the nucleons ( prefragments) present at the beginning of
 the minimization. 
After about 40 $\eta$ the results are
quite stable. For a smaller values of $ \eta $, 
there are fluctuations
in the results and in addition, SACA overestimates the size of 
A$^{max}$ and underestimates  consequently the IMF's production.
We fix the length of the markov chain $\eta$= 40. 
The effect of different M$_{ch}$'s on the evolution of 
the most bound configuration is shown in fig. 8 where 
the evolution of the fragment's multiplicity is plotted 
as a function of the iterations for two values of 
Markov's chains  i.e for  $M_{ch}$ = 40 $\eta$ and 450 $\eta$, respectively.
Note that the number of iterations is = $\eta$ times the number of temperature
steps.
We see that the initial evolution is quite the same in both cases, but a
smaller value of $M_{ch}$ needs less iterations than a longer one to
arrive at same final value. This
is easy to understand. The main aim of iterating over large number of
iterations with same $\vartheta$ is to establish the quasi-equilibrium. Once
an equilibrium is established, there will be no further improvements in the
cost at same $\vartheta$, therefore, a smaller value of $M_{ch}$ leads
to same result as that with largest $M_{ch}$.


\fbox{iv}. {\it Decrement in the Control Parameter :} 
The minimization is started with a relative large temperature $\vartheta_i$ 
. Then the temperature is decreased  
in steps  after a quasi-equilibrium is established for each
temperature. Apparently, a larger 
decrement in the temperature   
will lead the defect to be frozen i.e.  any configuration which may
or may not be the most bound can freeze whereas a very small decrement
will need huge computational time.  
The decrement should be in such a way that the
length of the Markov chain $M_{ch}$ is as small
as possible and thus after a new
change of control parameter $\vartheta$, the 
quasi-equilibrium should be re-established
as soon as possible. We here follow the simple rule for the decrement factor
$\sigma$.
\begin{equation}
\vartheta_{i+1} = \sigma \cdot \vartheta_i 
\end{equation}
The value of $\sigma$ varies in the literature between 0.5 to 0.95
  \cite{aarts87},\cite{kirk84},
  \cite{john87},\cite{nahar85}.
  The effect of different decrement factors $\sigma$ 
  is displayed in fig. 9. Here all other
 parameters are kept the same as discussed at the beginning. One can see that
 a very small value of $\sigma$ overestimates the size of $A^{max}$ and 
 underestimates the IMF production. 
 We fix the value of $\sigma$ to 0.85. 
 The comparison of two simulations resulting
 from the decrement factor $\sigma$= 0.85 and 0.98 is displayed in fig. 10.
 Here we see that the two different values gives the same cooling result
 but for the larger value of $\sigma$ many more iterations are necessary. 
\\

\fbox{iii}. {\it The Final Value $\vartheta_f$}:
The termination procedure used in the
literature varies from problem to problem and also from author to
author. We fix the termination by two different controls. 

1. Either we stop the calculations 
if the control parameter $\vartheta$ has reached a very small value where
no further transition can be  expected. For
the present calculation, we take $\vartheta_f$ = $10^{-10}$ MeV. 

2. Or we terminate the algorithm if there is no 
change in the configuration over a large number of attempted iterations..
Following the rule used to fix the length of the Markov chain, we choose
the length for termination  $l_{term}$ in terms of $\eta$ which represent the
number of  iterations in the units of $M_{ch}$. 
In fig. 11, we display the effect of the variation in $l_{term}$ on the fragment
distribution. We find that  different termination
lengths do not  affect the results. The  effect of l$_{term}$ on the 
evolution of the most bound configurations is shown in fig. 12 where the
evolution as a function of iterations is displayed for two values of
$l_{term}$. i.e for  
$l_{term}$ = 5 $\eta$ and 120 $\eta$, respectively. The 
different termination values have a very small effect on the fragment
structure. Therefore, we fix $l_{term}$ to 35 $\eta$. 
\\
In above paragraphs, we have discussed in detail the influence of different
choices of the 
parameters which  determine the simulated annealing method. One should
note that once these parameters  are chosen, the simulated annealing method
is  completely determined and it is a complete self-iterative method. 
\\

In the further discussion, the
set with  conservative parameters  ( i.e. with $\vartheta_i$ = 500 MeV, 
$M_{ch} = 450 \eta$, $\sigma= 0.98$ and  $l_{term}= 120 \eta$) is called  as
S$_l$ whereas the set with the most economical 
parameters ( i.e. with $\vartheta_i=5 $ MeV,
$M_{ch}= 40 \eta$, $\sigma= 0.85$ and $l_{term}$ = 35 $\eta$) is called as
S$_{ec}$. \\

The crucial test of the algorithm
is its application to a single nucleus in its ground state.  In principle
one should get a single nucleus at the end. But the results can be different
in reality. To see the importance of nucleon and fragment exchange
procedured in SACA, we first turned off the fragment exchange part of the
algorithm ( i.e. 2nd stage of the algorithm). As expected, the transfer
of single nucleons terminates in local minima and as a result, one finds
several fragments in the ground state of a nucleus. Naturally, the global
minima is a single nucleus. The energy gain in a single nucleon transfer
is not enough to overcome the huge energy barrier. This energy barrier 
can be overcome only by allowing the collective transfer of the nucleons 
i.e. of fragment as such. When we turned on the fragment exchange procedure
of the algorithm, we could overcome the local minima and we find single
nucleus as most bound configuration.
In fig. 13, we show the
evolution of the fragments as a function of the iterations for different single
nuclei $ ^{20}$Ne, $^{40}$Ca, $^{93}$Nb  and $ ^{208}$Pb, respectively using
$S_{ec}$. Starting points are the (ground) state nuclei as generated by the 
QMD.
 In all cases, the SACA finds the single nucleus at the end
 of the iterations as it should.
 One also notices that the lighter nuclei need less
 iterations to find the most bound configuration. We find $\approx$
 $
  8,000, 12,000, 62,000$ and $280,000$  iterations are  necessary
  to find the ground state for $^{20}Ne,
 ^{40}Ca, ^{93}Nb $ and $ ^{208}Pb$, respectively.  One also notices that
 the increase in the number of 
 necessary iterations is not a linear function of the masses
 of nuclei. The energy of the  configurations is displayed in fig.14.
 We notice that one has a positive energy at the beginning which is decreased
 by breaking
 the cells into large number of nucleons/fragments. 
 After a large number of  iterations,
 one finally reaches the most bound configuration. \\
\\
In fig. 15, we display the multiplicity
(averaged over 20 events) of different fragments obtained using $S_l$ and
$S_{ce}$, respectively. Here Au-Au collision is carried out at
impact parameter of 8 fm. We see that both sets of parameters give a similar
evolution of the reaction. One should note that the minimization with
$S_{ec}$ needs much less computing time as compared to $S_l$. 
 Our algorithm is able to detect  the
 fragment distribution as early as 50-60 fm/c. From fig. 1, one notices that
 the density is maximum at this time. This very early
 identification of fragments in SACA is very promising
 because it means that the fragments may give insight into hot and dense
 nuclear matter.\\

The annealing algorithm can be made faster if some pre-information is
feed into the algorithm. 
Naturally, the nucleons which are very far away in spatial or 
in momentum space will not lower the energy if  one combines them as 
fragment. We applied
a cut in  spatial and in momentum space to sort out those distant 
nucleons. 
We took  a minimal spatial distance between two
nucleons of 10 fm and  a relative momentum of 200 MeV/c. In other words,
we first break the whole system into fragments using these conditions and
each of these fragments are then subjected to SACA. We found that the results
are the same as before but the algorithm is about 10 times more faster.\\ 
\\
\begin{center}
{\huge IV. Summary:}
\end{center}
\vspace*{0.2cm}

Summarizing, based on the simulated annealing method,
we have presented the details
of a new algorithm developed to study multifragmentation 
of heavy ion collisions. We have carried out an extensive survey of the
different parameters which are crucial for the success of the method. Based
on our calculations, a set of parameters is suggested for 
the algorithm which makes the algorithm very fast and accurate. This 
algorithm can detect the fragments as early as 40-60 fm/c i.e. at time
when density is relative high and the interactions between fragments 
are still going on. 
It can not only give insight into the hot and dense nuclear matter,
but at the same time makes it possible to apply the full in-medium
G-matrix approach \cite{bohn89} 
to study the multifragmentation which was not possible
due to the emission of nucleons after 70-80 fm/c \cite{Rami95}.
A brief outcome of the results was presented in ref. \cite{puri96} and a 
detailed physical interpretation of our results for various reactions will be
presented elsewhere \cite{Puri98}. The algorithm is very general and may 
serve for every problem in which the most bound configuration has to be 
found.
\vspace*{0.2cm}
\\

{\it One of us (RKP) appreciates the warm hospitality of SUBATECH, Ecole
des Mines de Nantes, Nantes, France where this work was done. This work
is supported by the CNRS and Ministry of Industry, Government of France.
}\newpage
{\huge Figure Captions:}\\

{\bf Fig. 1}. Evolution of  Au-Au collisions at incident energy of 600 MeV/
nucl.  using a soft equation of state. 
The results at  b = 3 and 8 fm are displayed, respectively,   by
dotted and solid lines. Fig. 1(a) displays the mean density whereas the
rate of collision is shown in Fig. 1(b). The evolution of the spectators (
filled circle) and the 
participants (filled triangle) is are presented in Fig. 1(c).
Fig. 1(d) shows the time evolution of the transverse flow of the nucleons. Here, 
we do not consider the formation of fragments. The evolution of the largest
mass $A^{max}$ formed within MST and MST with binding energy check (MST$^\star$)
are
displayed in fig. 1(e) and Fig. 1(f), respectively.\\

{\bf Fig. 2}. The heaviest fragmnent $A^{max}$, the 
emitted  nucleons, the multiplicity
of fragments with mass A $\ge$ 5 and the total energy associated with the
 configuration is displayed as a function of the cell number. Here the results
at 0 , 40 and 200 fm/c are represented, respectively, by filled circle,
open square and filled triangle. The displayed results are for a single 
event  generated using QMD model.\\

{\bf Fig. 3}.  The evolution of the most bound configuration 
as a function of the 
iterations. Here we display the results at four times i.e.
at 0, 40, 120 and 200 fm/c, respectively.  The solid and dotted lines
represents the results obtained with cells = 2 and and 394, 
respectively.\\

{\bf Fig. 4} Same as fig. 2, but as a function of
Monte -Carlo procedures.\\

{\bf Fig. 5} Same as fig. 4, but as a function of the initial
temperature $\vartheta_i$. Here the arrow shows the value of the parameter
chosen for the optimized set of parameters.\\

{\bf Fig. 6} Same as fig. 3, but with temperature $\vartheta_i$= 5 MeV
( solid line) and 500 MeV ( dotted line), respectively.\\

{\bf Fig. 7} Same as fig. 5, but as a function of the length of the
Markov chain $M_{ch}$.\\

{\bf Fig. 8} Same as fig.3, but with $M_{ch}= 40 \eta$ and $ 450 \eta$,
respectively.\\

{\bf Fig. 9} Same as fig. 5, but as a function of the decrement factor
$\sigma$.\\

{\bf Fig. 10} Same as fig. 3, but with $\sigma$ = 0.85 and 0.98,
respectively.\\

{\bf Fig. 11} Same as fig. 5, but as a function of 
termination length $l_{term}$.\\

{\bf Fig. 12} Same as fig. 3, but with $l_{term}$ = $ 35 \eta $ and$ 120 \eta$.\\

{\bf Fig. 13} Same as fig. 3, but the evolution of single nuclei
$ Ne, Ca, Nb $ and $ Pb$, respectively.\\

{\bf Fig. 14} Same as fig. 13, but the energy of the system as a function
of the iterations.\\

{\bf Fig. 15} The time evolution of the Au-Au at 600 MeV/nucl and at an
impact parameter of 8 fm using a soft equation of the state. Here we
display the results which are averaged over 20 events.  The results obtained
using $S_l$ and $S_{ec}$ are shown , respectively, by the 
filled circles and open squares, respectively.
\newpage 

\begin{figure}[h]
\epsfxsize=11.cm
$$
\epsfbox{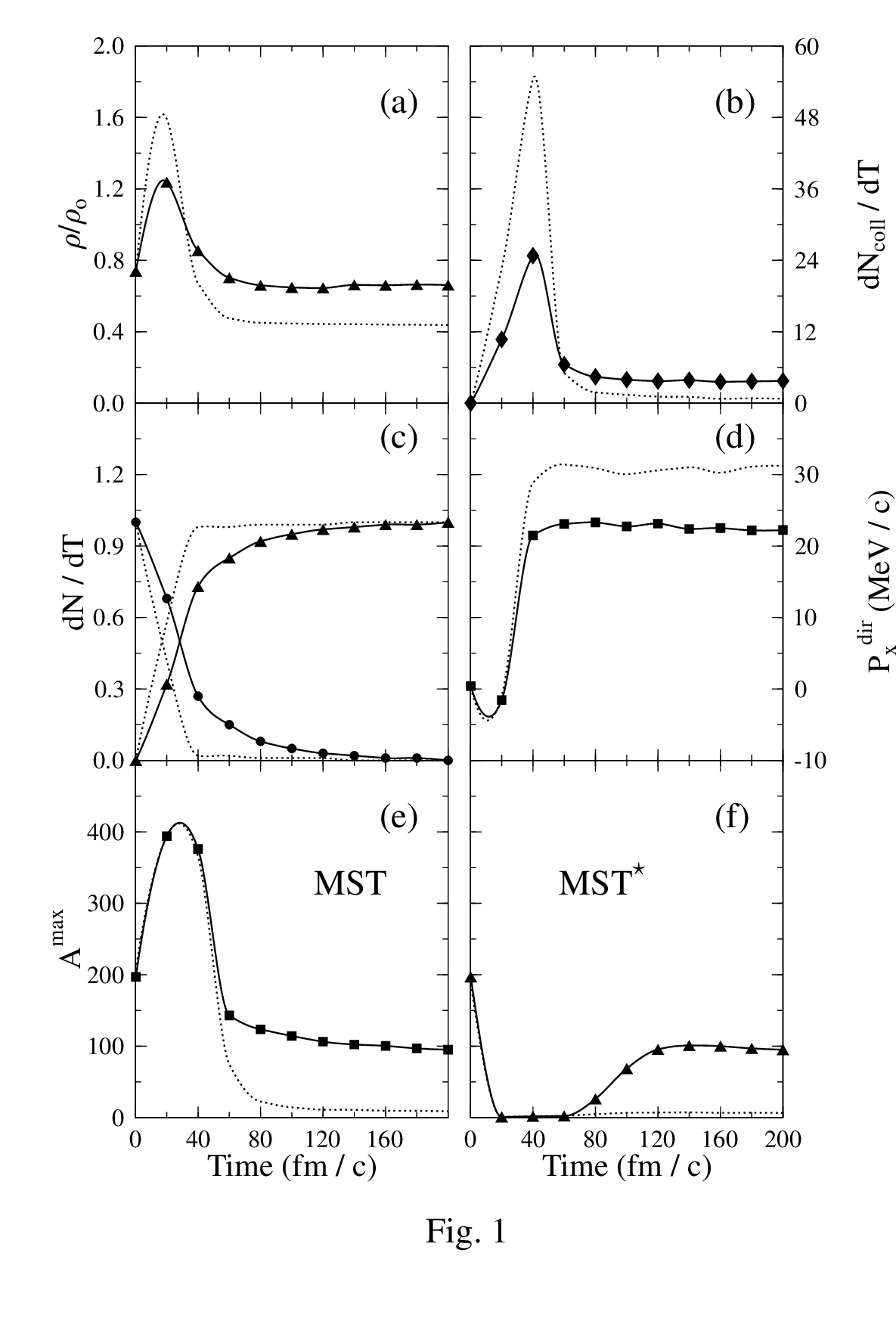}
$$
\end{figure}
\begin{figure}[h]
\epsfxsize=11.cm
$$
\epsfbox{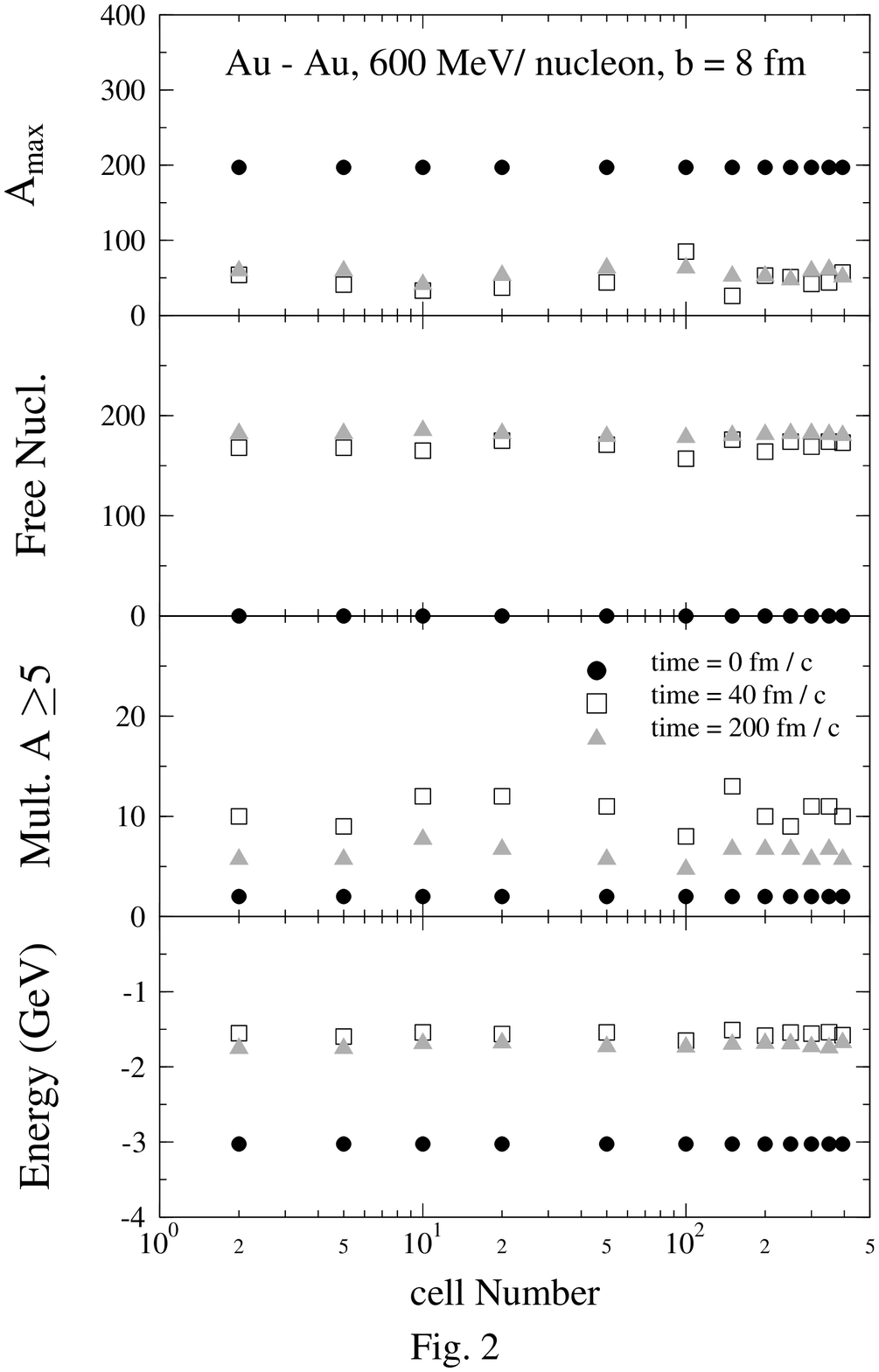}
$$
\end{figure}

\begin{figure}[h]
\epsfxsize=11.cm
$$
\epsfbox{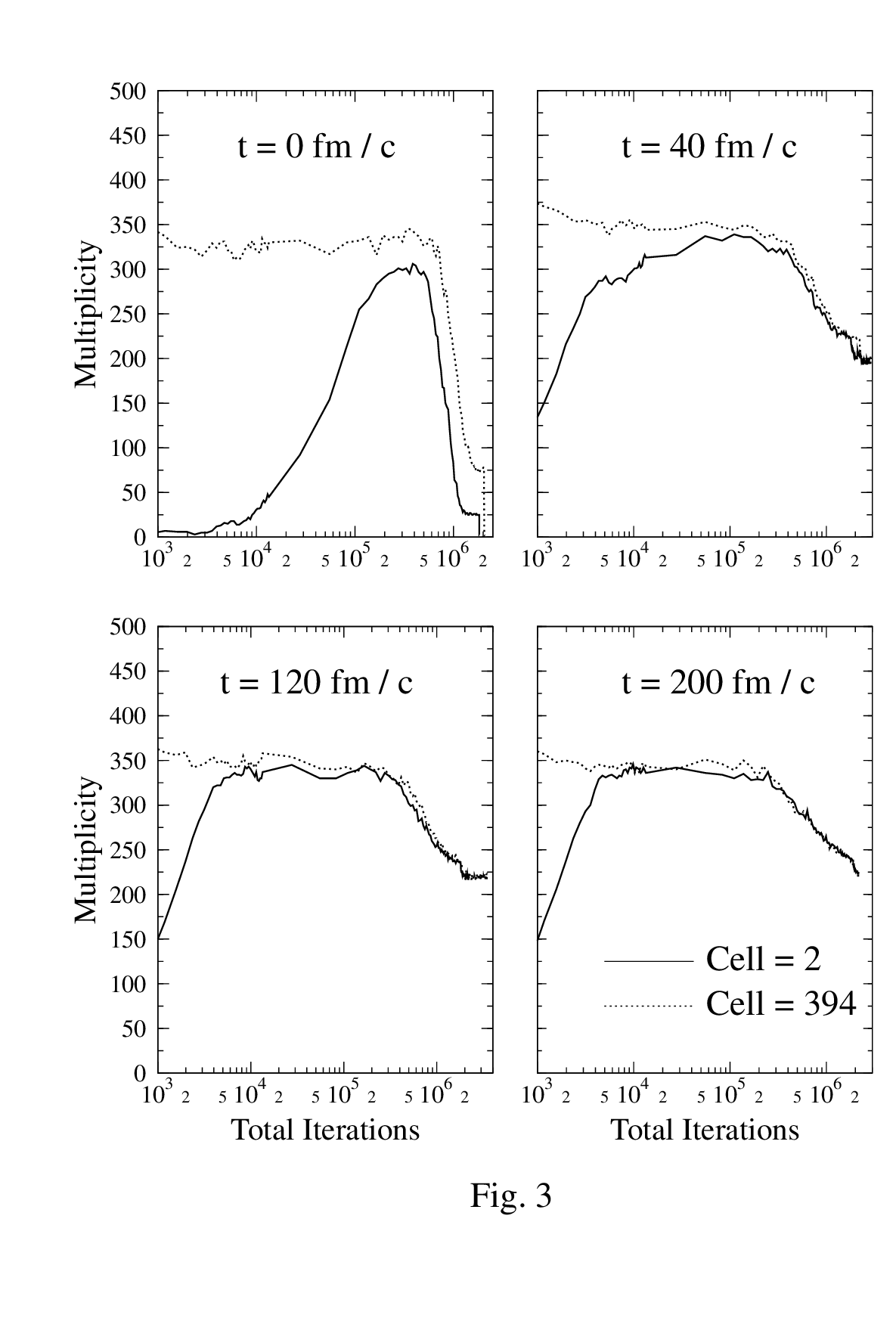}
$$
\end{figure}

\begin{figure}[h]
\epsfxsize=11.cm
$$
\epsfbox{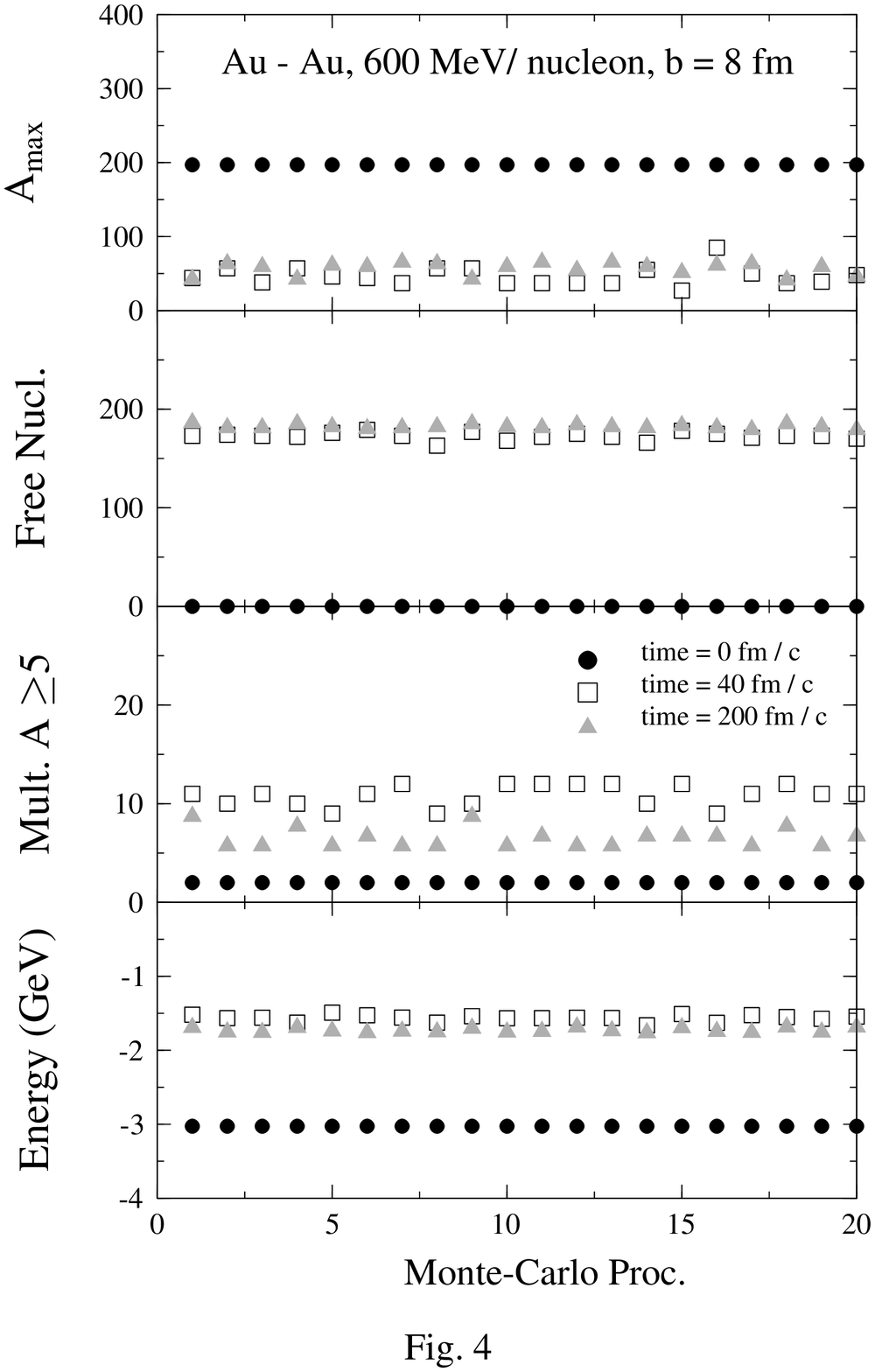}
$$
\end{figure}

\begin{figure}[h]
\epsfxsize=11.cm
$$
\epsfbox{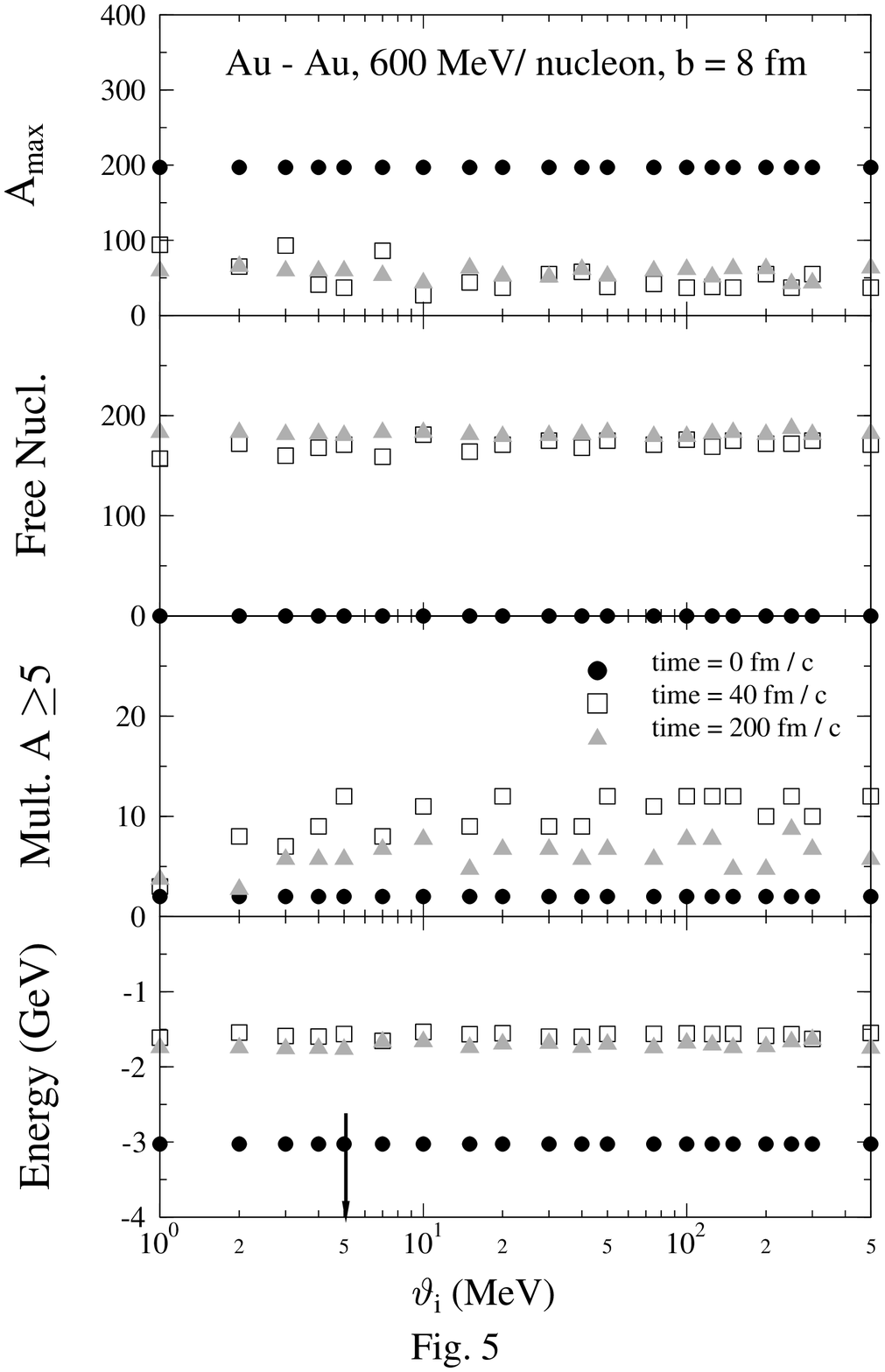}
$$
\end{figure}

\begin{figure}[h]
\epsfxsize=11.cm
$$
\epsfbox{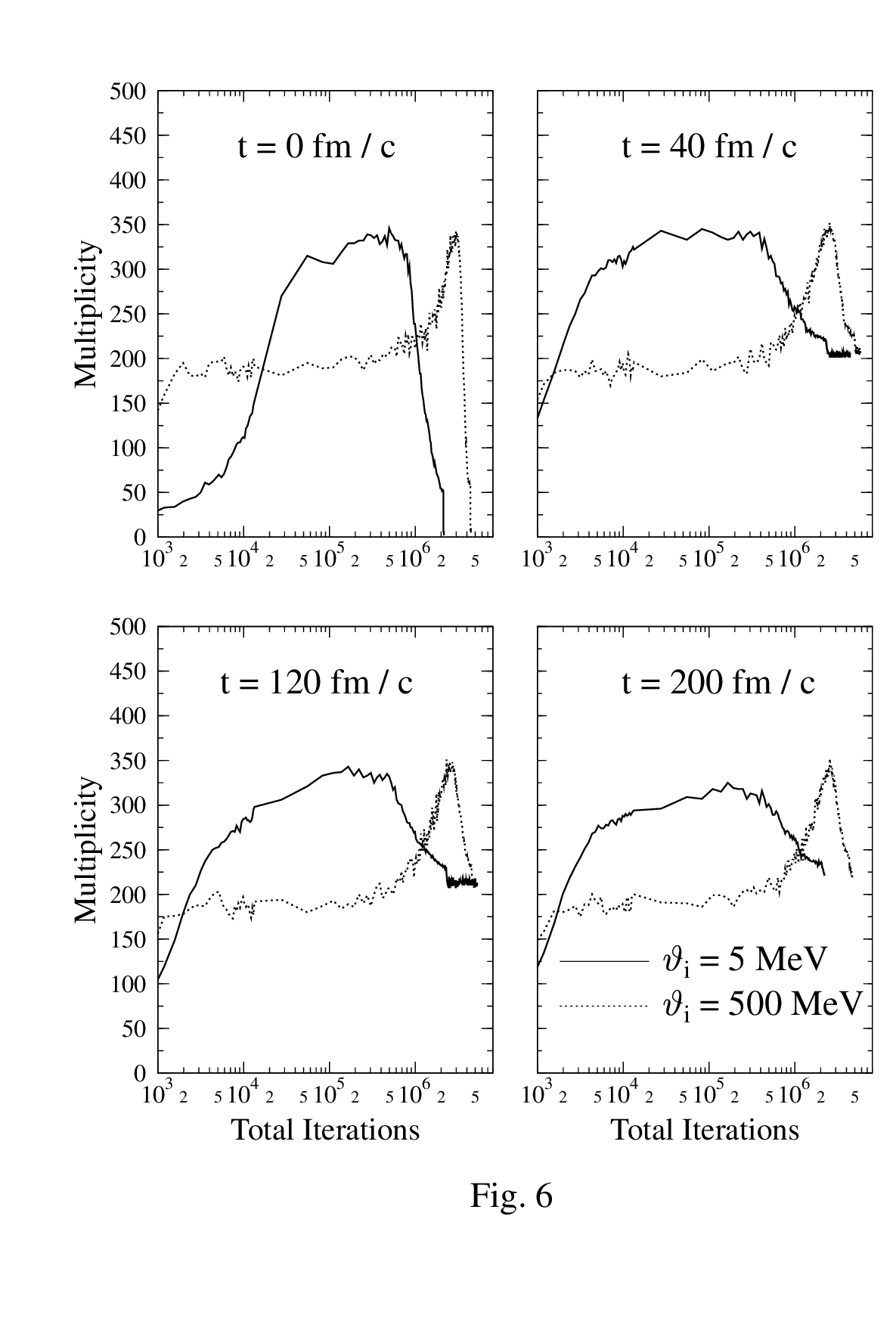}
$$
\end{figure}

\begin{figure}[h]
\epsfxsize=11.cm
$$
\epsfbox{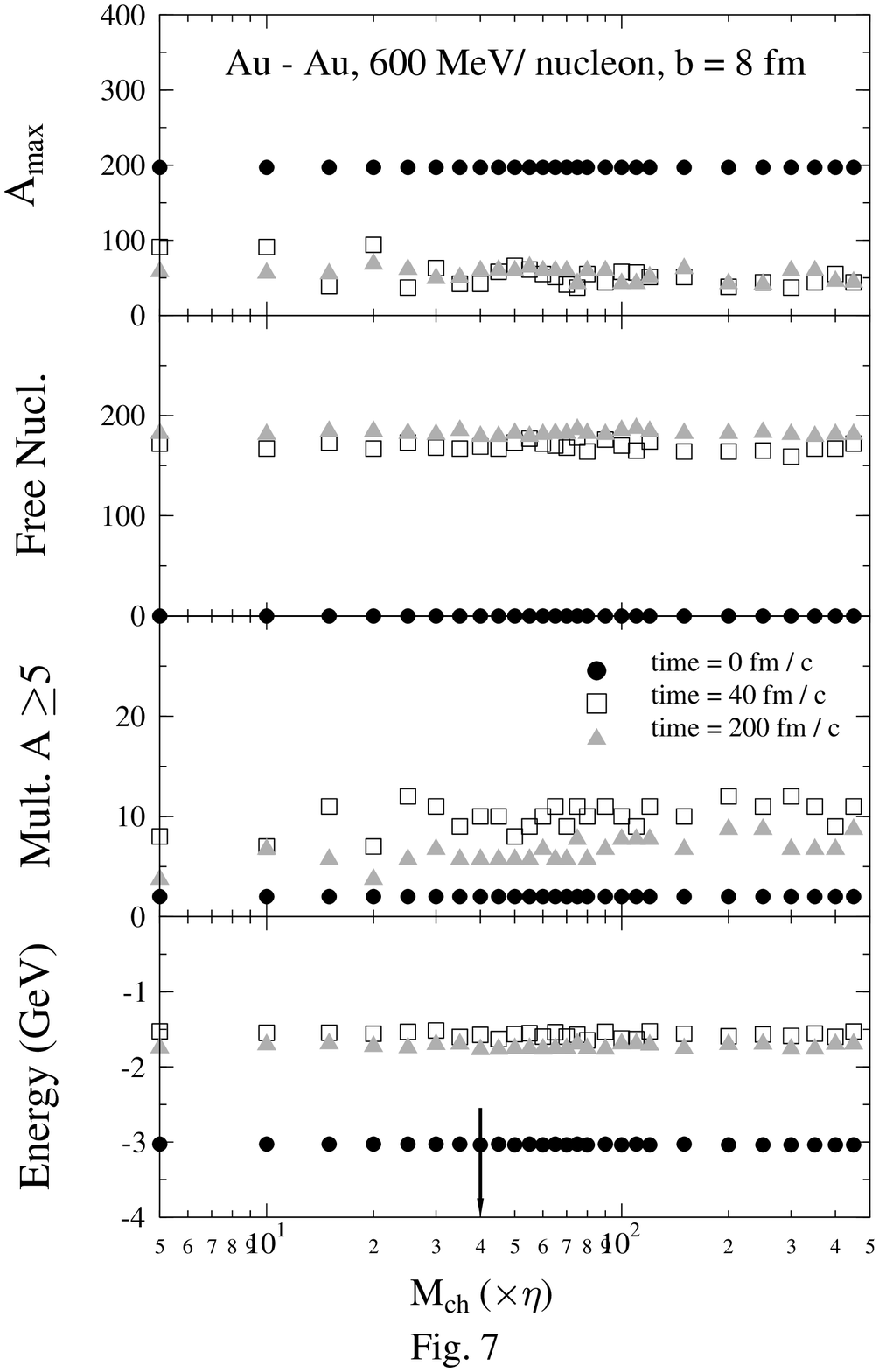}
$$
\end{figure}

\begin{figure}[h]
\epsfxsize=11.cm
$$
\epsfbox{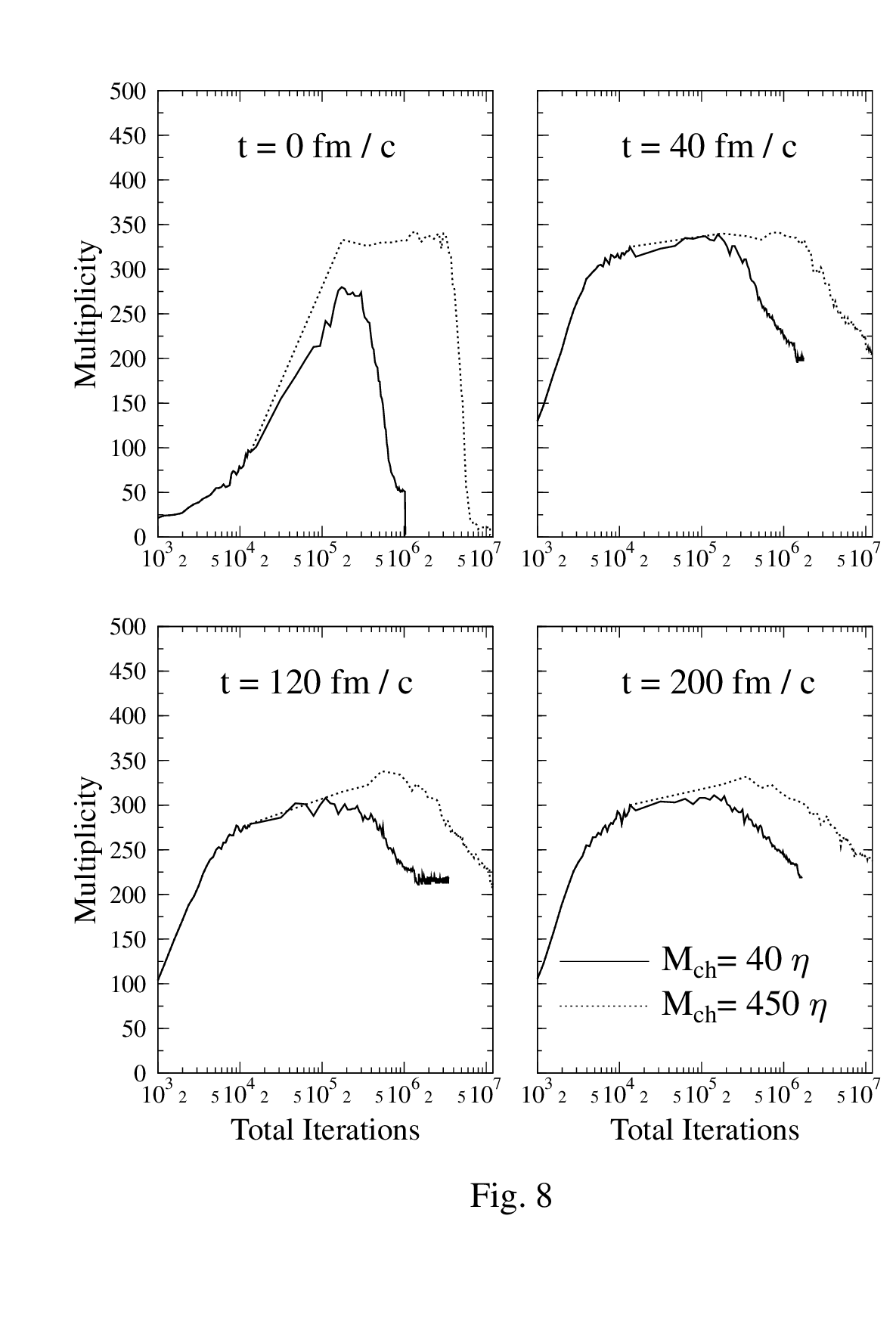}
$$
\end{figure}

\begin{figure}[h]
\epsfxsize=11.cm
$$
\epsfbox{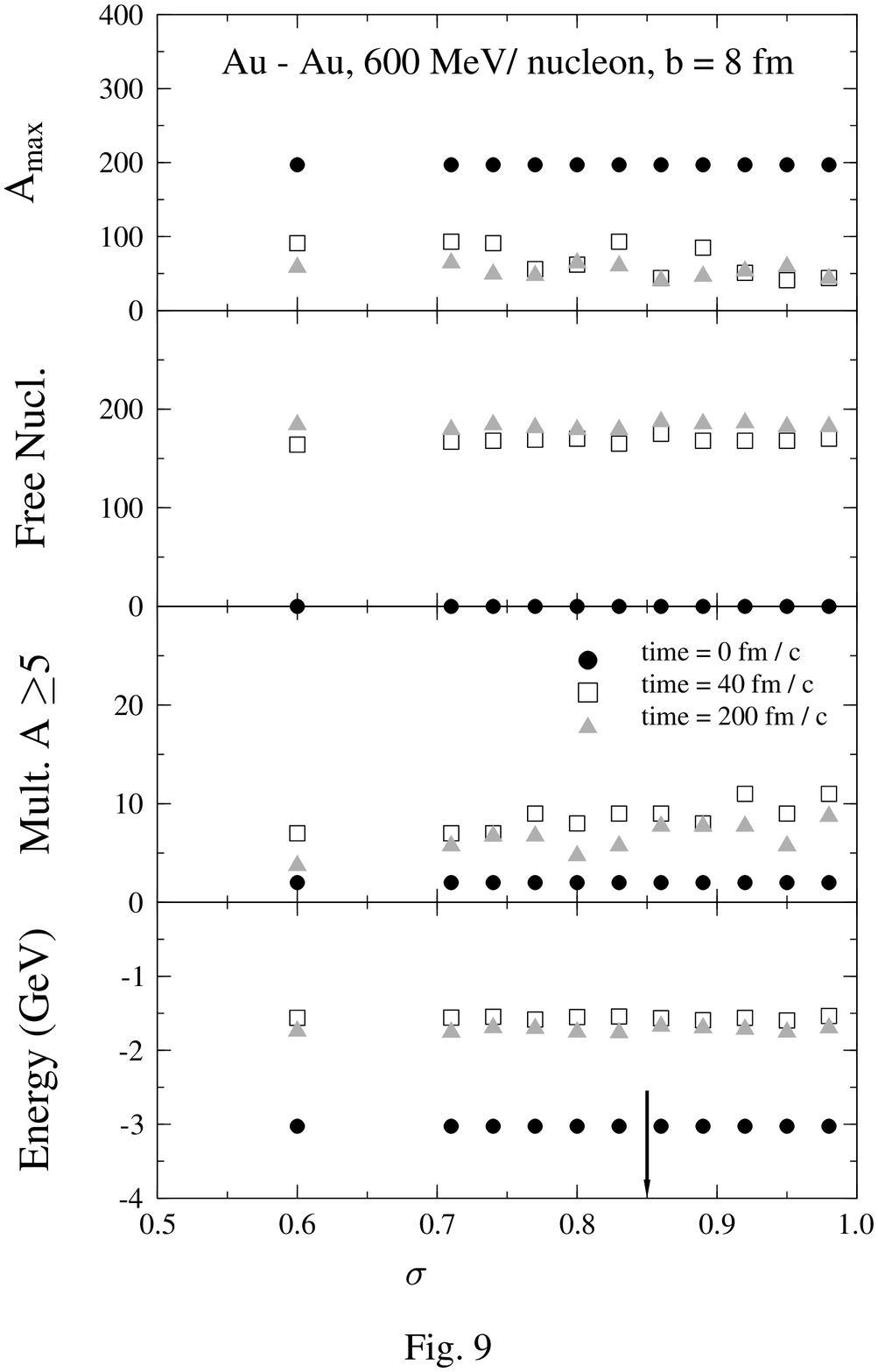}
$$
\end{figure}

\begin{figure}[h]
\epsfxsize=11.cm
$$
\epsfbox{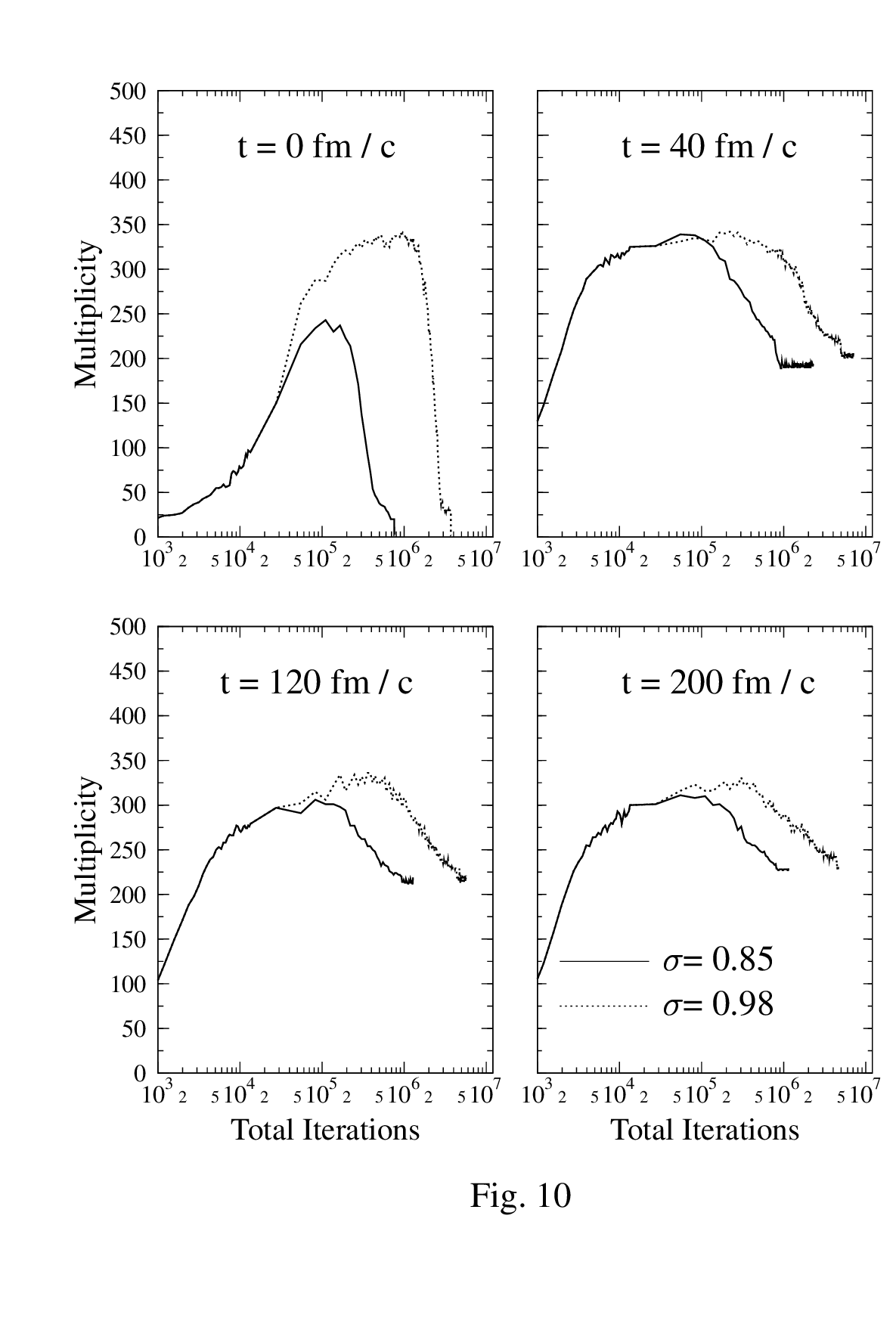}
$$
\end{figure}

\begin{figure}[h]
\epsfxsize=11.cm
$$
\epsfbox{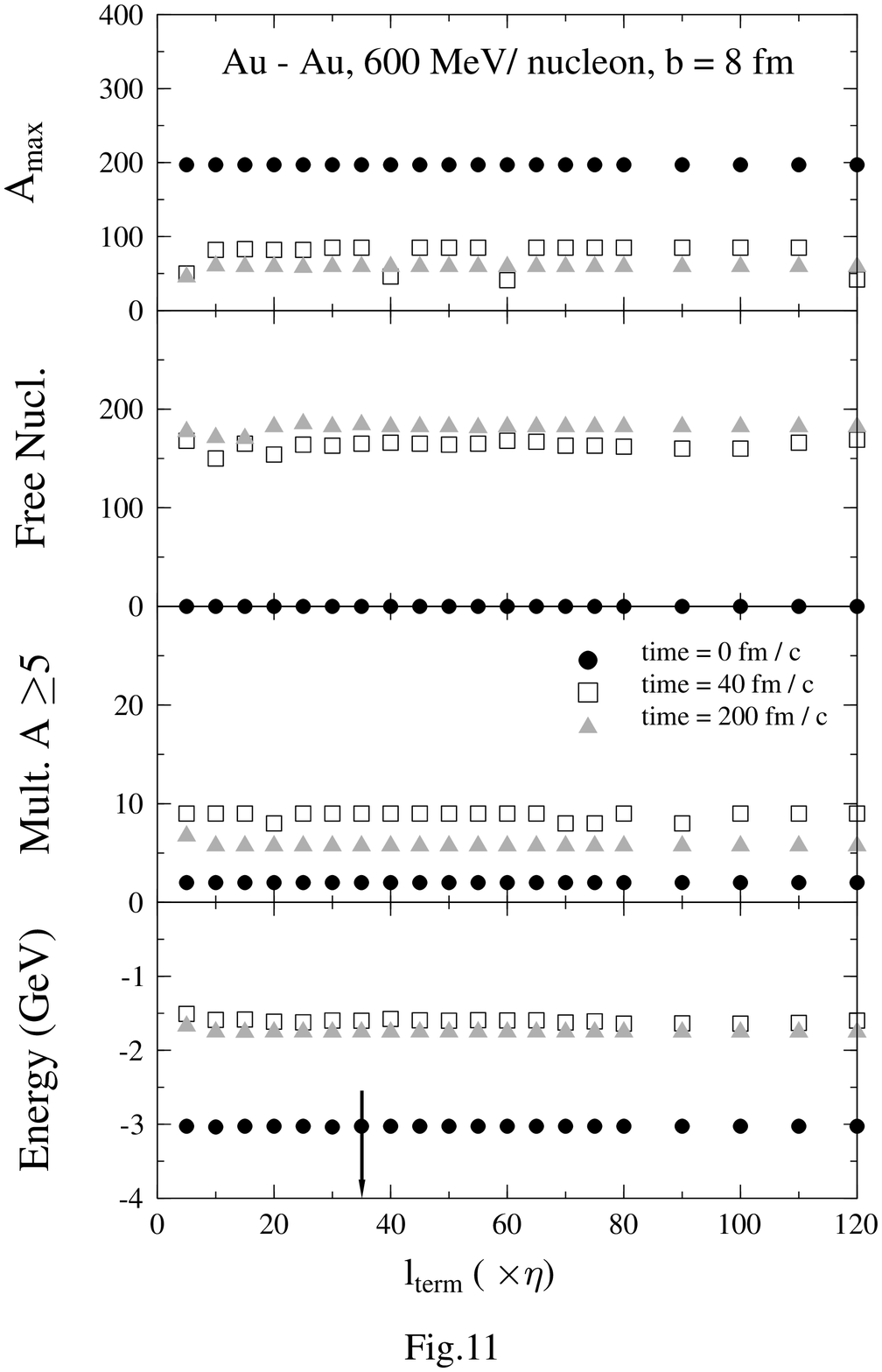}
$$
\end{figure}

\begin{figure}[h]
\epsfxsize=11.cm
$$
\epsfbox{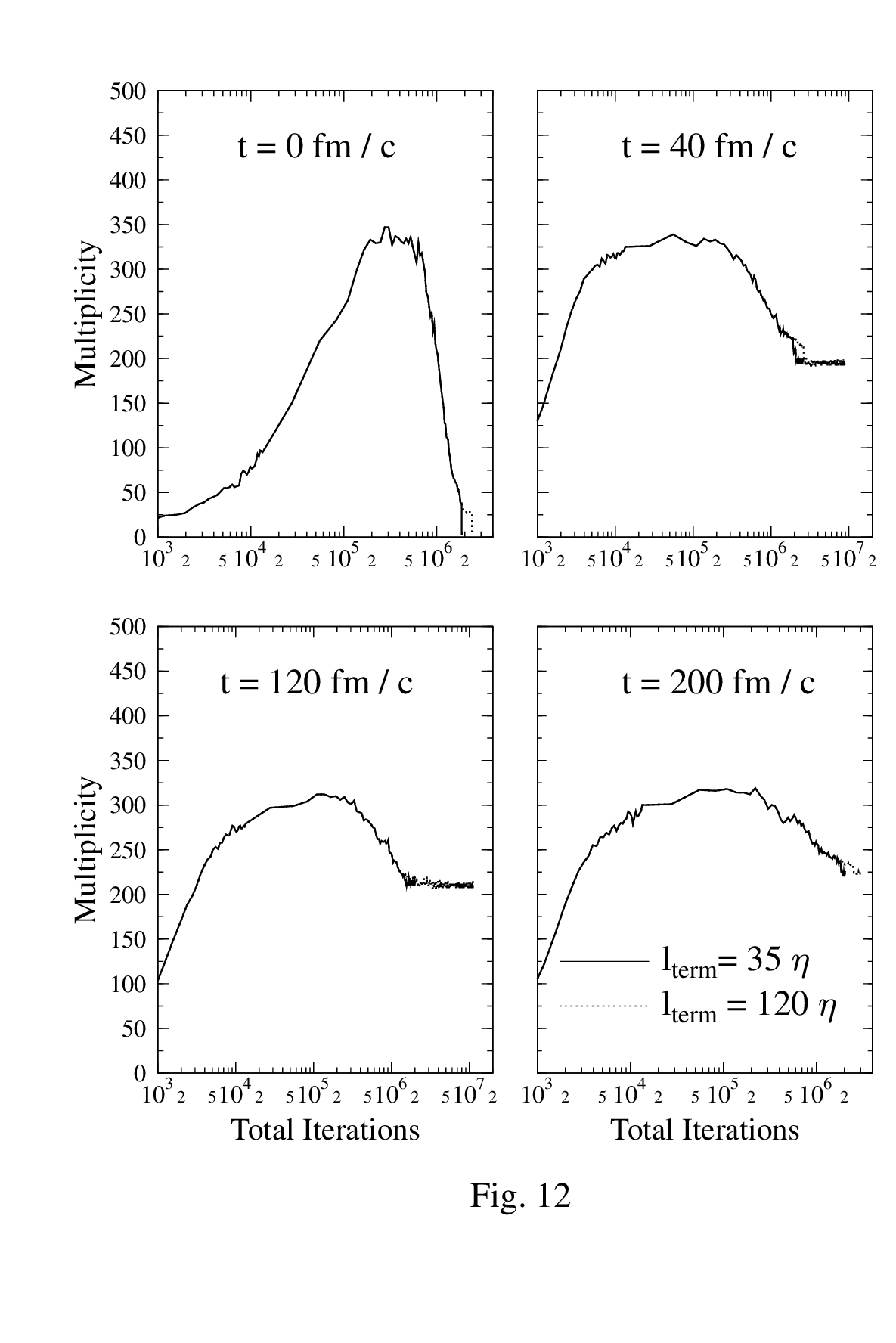}
$$
\end{figure}

\begin{figure}[h]
\epsfxsize=11.cm
$$
\epsfbox{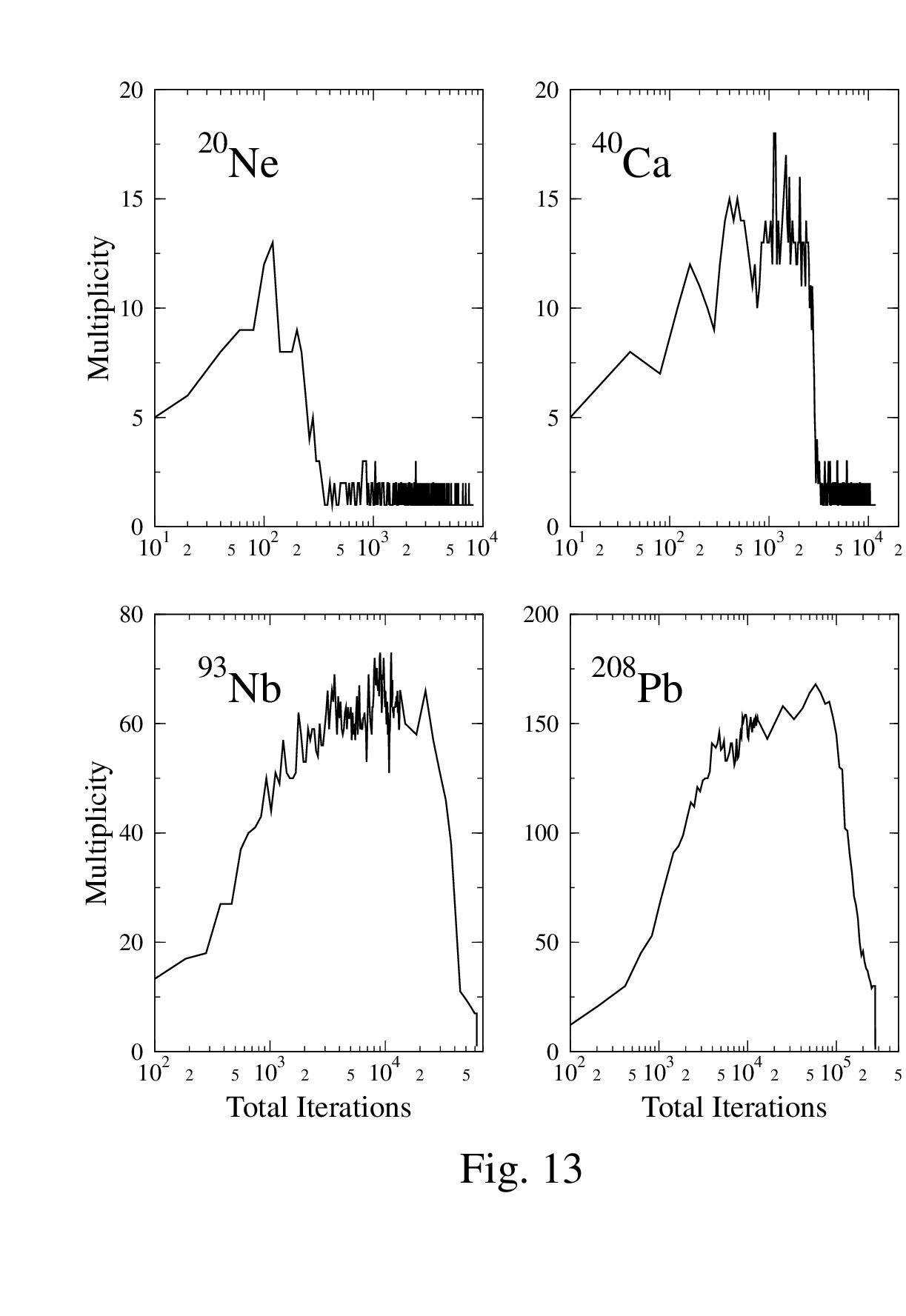}
$$
\end{figure}

\begin{figure}[h]
\epsfxsize=11.cm
$$
\epsfbox{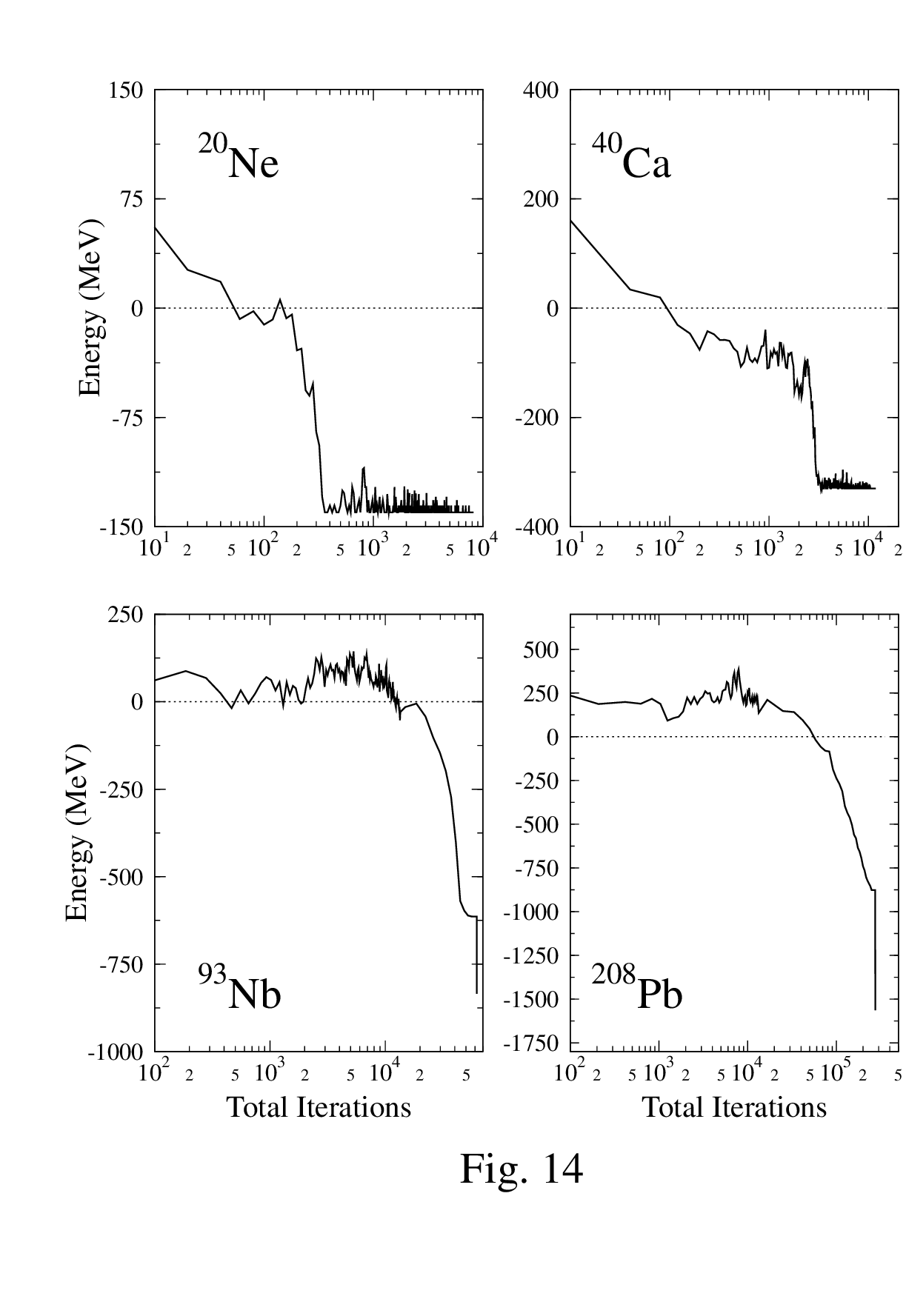}
$$
\end{figure}

\begin{figure}[h]
\epsfxsize=11.cm
$$
\epsfbox{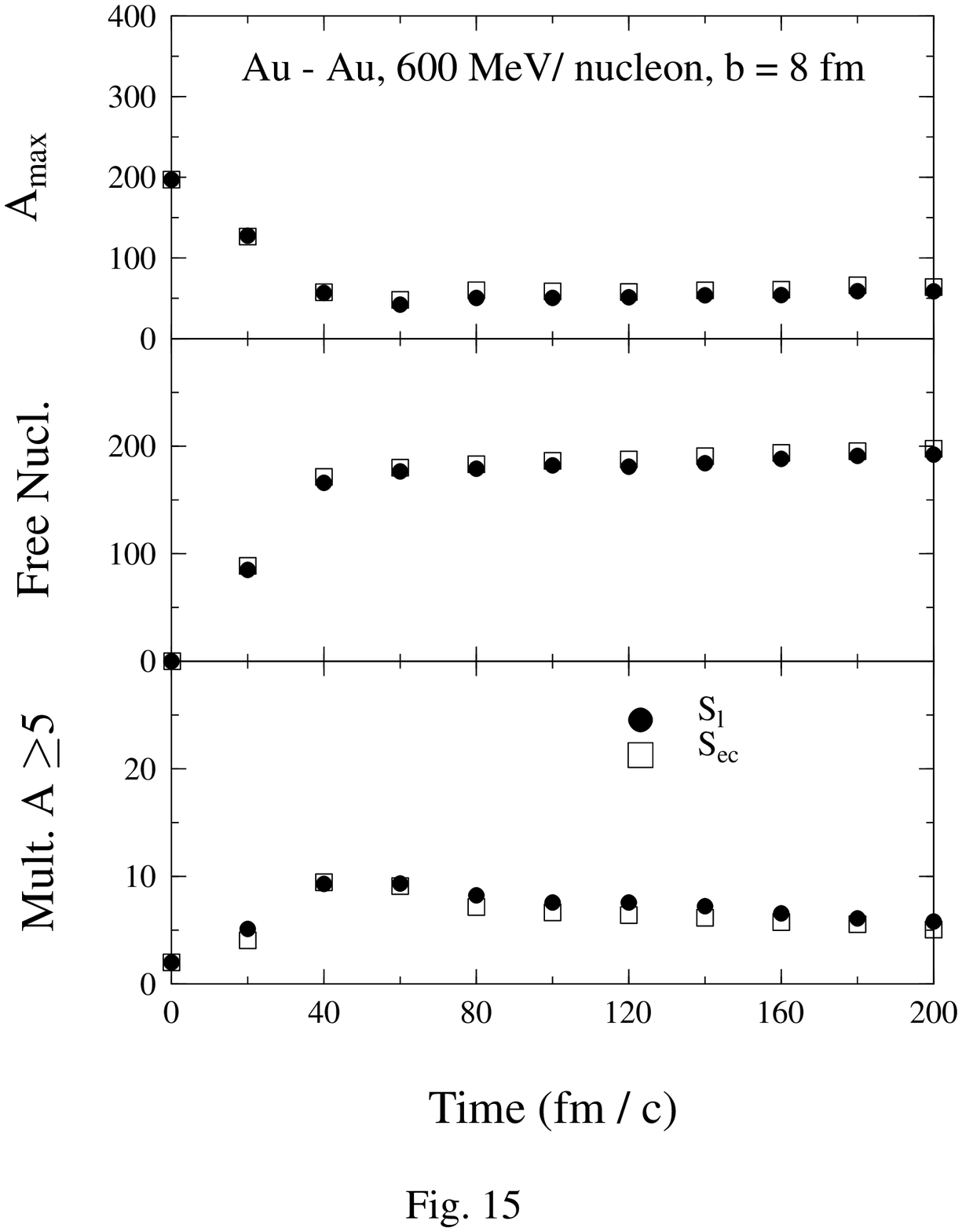}
$$
\end{figure}

\end{document}